\documentclass[namedreferences]{solarphysics}

\usepackage[hyperref,optionalrh]{spr-sola-addons} 
\usepackage{graphicx}        
\usepackage{color}           

\newcommand{\kms}{\,km\,s$^{-1}$}

\newcommand{\degr}{{\hbox{$^\circ$}}}

\newcommand{\arcsec}{{\hbox{$^{\prime\prime}$}}}

\newcommand{\Mg}{Mg{\sc\,ii}}
\newcommand{\Cb}{C{\sc\,ii}}
\newcommand{\Si}{Si{\sc\,iv}}

\newcommand{\be}{\begin{equation}}
\newcommand{\ee}{\end{equation}}
\newcommand{\bea}{\begin{eqnarray}}
\newcommand{\eea}{\end{eqnarray}}
\newcommand{\beas}{\begin{eqnarray*}}
\newcommand{\eeas}{\end{eqnarray*}}

\chardef\us=`\_ 

\begin{document}

\begin{article}

\begin{opening}

\title{Explosive Events in the Quiet Sun Near and Beyond the Solar Limb Observed with the {\it Interface Region Imaging Spectrograph} (IRIS)}

\author[addressref={aff1},corref,email={calissan@uoi.gr}]{\inits{C.E.}\fnm{C. E.}~\lnm{Alissandrakis}} 
\author[addressref={aff2}]{\inits{J.-C.}\fnm{J.-C.}~\lnm{Vial}} 

\address[id=aff1]{Section of Astro-Geophysics, Department of Physics, University of Ioannina, GR-45110 Ioannina, Greece}
\address[id=aff2]{Institut d'Astrophysique Spatiale, CNRS (UMR 8617) Universit\' e Paris-Saclay, Orsay,  France} 

\runningauthor{C.E. Alissandrakis, J.C. Vial}
\runningtitle{Explosive Events Near and Beyond the Solar Limb}

\begin{abstract}
We study {point-like} explosive events (EE), characterized by emission in the far wings of spectral lines, in a quiet region near the South Pole, using {\it Interface Region Imaging Spectrograph} (IRIS) spectra at two slit positions, slit-jaw (SJ) observations and Atmospheric Imaging Assembly (AIA) images. The events were best visible in \Si\ spectra; they were weak in SJs, occasionally visible in 1600\,\AA\ and 304\,\AA\ AIA images, and invisible in higher temperature AIA images. We identified EEs from position--time images in the far wings of the \Si\ lines and measured their distance from the limb. A Gaussian model of the height distribution showed that EEs occur in a narrow (0.9\arcsec) height range, centered at 3.2\arcsec\ above the continuum limb at 2832.0\,\AA. On the disk, we found that they occur in network boundaries. Further, we studied the line profiles of two bright EEs above the limb and one on the disk. We found that what appears as broad-band emission, is actually a superposition of 2\,--\,3 narrow-band Gaussian components with well-separated line profiles, indicating that  material is expelled towards and/or away from the observer in discrete episodes in time and in space. The expelled plasma accelerates quickly, reaching line-of-sight (LOS) velocities up to 90\kms. {Overall, the motion was practically along the LOS, as the velocity on the plane of sky was small.} In some cases tilted spectra were observed that could be interpreted in terms of rotating motions of up to 30\kms.We did not find any strong absorption features in the wing of the \Si\ lines, although in one case a very weak absorption feature was detected. No motions, indicative of jets, were detected in SJ or AIA images. Reconnection in an asymmetric magnetic-field geometry, in the middle or near the top of small loops is a plausible explanation of their {observational characteristics}.
\end{abstract}
\end{opening}

\section{Introduction} \label{intro} 
{Short duration, small}-scale brightenings  are ubiquitous in the quiet Sun, readily observed in spectral lines and continua in the microwave, mm, UV, EUV, and X-ray wavelength ranges (e.g. \citealp{1973ApJ...185L..47V}; \citealp{1981SoPh...69...77H}; \citealp{1983ApJ...272..329B}; \citealp{1994ApJ...431L.155K}; \citealp{1997SoPh..175..341I}; \citealp{1999A&A...341..286B}; \citealp{2001SoPh..198..313D}; \citealp{2014Sci...346C.315P}; \citealp{2015ApJ...813...86I}; \citealp {2020A&A...638A..62N}; \citealp{2021ApJ...914...70J}). They are known by a large variety of names: {\it {coronal} bright points\/} {(\citealp{1973ApJ...185L..47V}; for a review see \citealp{2019LRSP...16....2M})}, {\it brightenings\/}, {\it transient phenomena\/}, {\it explosive events\/} {(\citealp{1983ApJ...272..329B})}, {\it miniflares\/}, {\it microflares\/}, {\it nanoflares\/}, {\it UV bursts\/} {(for a review see \citealp{2018SSRv..214..120Y})}, {\it IRIS bombs\/} {(\citealp{2014Sci...346C.315P})}, {\it campfires\/} (\citealp{2021A&A...656L...4B}), and so on. { However}, it is not clear that all these names {correspond to different phenomena or are manifestations of similar phenomena observed with instruments of different capabilities and/or in different atmospheric layers. It is even less clear whether they involve different physical processes, as they are all commonly attributed to magnetic reconnection.}
 
{These phenomena} are characterized by their small size (a few arcseconds) and short duration (a few minutes) and span a large energy range, from that of small flares down to that of the hypothetical nanoflares proposed by \cite{1988ApJ...330..474P}, {and} are suspected of having a bearing on the heating of the chromosphere and the corona (see, e.g., \citealp{2006SoPh..234...41K}; \citealp{2019ARA&A..57..189C}; also Section 7 in \citealp{2020FrASS...7...74A}). 
Here we will use the term {\it explosive events} (EE), {in its generic sense}, to emphasize both their transient and their explosive (rather than gradual rise and fall) nature. {This does not imply that we consider all reported events to be of the same nature, as a detailed examination of this issue is outside the scope of the present work.}

The {\it Interface Region Imaging Spectrograph} (IRIS: \citealp{2014SoPh..289.2733D}) has provided a wealth of information on the chromosphere--corona transition region (TR), including events with very broad line profiles and with absorption components often superposed (e.g. \citealp{2014Sci...346C.315P}; {\citealp{2014ApJ...797...88H};} \citealp{2016ApJ...824...96T};  \citealp{2017A&A...603A..95A}; \citealp{2018SSRv..214..120Y}; \citealp{2019ApJ...873...79C}{; \citealp{2022A&A...660A..45M}}). \cite{2019ApJ...873...79C} reported EEs on the solar disk, mostly associated with quiet-Sun network jets and classified them in four groups, according to their line profiles: enhanced in both wings, with two peaks of comparable intensity, enhanced in the blue wing, and enhanced in the red wing. They interpreted them in terms of bi-directional reconnection with upward and/or downward flows.

In spite of the extensive work on {such phenomena}, we have little observational evidence on their height in the solar atmosphere. In a recent work, \cite{2022A&A...657A.132K} made an extensive analysis of {events with broad} line profiles observed by IRIS, but they excluded events that did not contain absorption blends in \Si\ and did not find any events in pure quiet-Sun observations {or above the limb}. \cite{2015SoPh..290.2871T} reported an {event} beyond the limb, at a height of about 1 Mm. More recently, \cite{2022SoPh..297...76T} studied two IRIS ``sit and stare" data sets near the limb and reported that some spectral-line profiles showed enhancements in blue or red wings, while others showed enhancements in both wings. Indirect evidence about the height was provided by \cite{2014Sci...346C.315P}, who suggested that {their events} formed deep in the atmosphere, based on the absorption features, which they interpreted in terms of absorption by cold material; this interpretation was contested by \cite{2015ApJ...808..116J}, who pointed out that UV radiation cannot easily escape from the photosphere and suggested that {these events} form in the low\,-\,mid chromosphere or above. A more realistic interpretation of the absorption features was presented by \cite{2017A&A...603A..95A} for a two-loop interaction event; they attributed these features to absorption by a higher-lying, descending loop, approaching the already activated lower-lying loop in the process of their interaction.

In a previous work (\citealp{2018SoPh..293...20A}, hereafter Article I),  we analyzed IRIS spectral and slit-jaw observations of a quiet region near the South Pole. We provided average profiles of strong and weak spectral lines, radial variations of bulk spectral parameters, as well as formation heights. We also compared the \Mg\ k and h line profiles with non-LTE model computations and estimated values of the physical parameters. In this work we use the same data set to investigate the properties of EEs near and beyond the limb. In Section~\ref{obs} we provide a short description of the observations and in Section~\ref{overview} we present an overview of an event beyond the limb. Our measurements of the position of EEs and the derived distribution with height are presented in Section~\ref{height} and our detailed analysis of three events in Section~\ref{details}. We study the temporal evolution of EEs in Section~\ref{TimEvol} and we summarize our findings and discuss our conclusions in Section~\ref{summary}.

\section{Observations and Data Reduction} \label{obs}

The IRIS observations we used are described in detail in Article I. They were obtained near the South Pole on 24 February 2014, from 11:05 to 12:45 UT (OBSID 3800259459), in a region where no coronal hole was present. The spectrograph slit was oriented in the NS (north--south) direction, about 8\arcsec\ west of the South Pole. Every 10 seconds the position of the slit alternated between two locations, 2\arcsec\ apart; we will refer to the position nearest to the Pole as {\it position 1} and to the other slit position as {\it position 2}. With a cadence of 19 seconds, 318 spectra were taken at each slit position during the 100 minute long observing sequence. The slit extended from $\approx50$\arcsec\ inside the disk ($\mu=0.32$) to $\approx65$\arcsec\ beyond the limb, providing a good coverage of structures both near and above the limb.

Spectra were recorded in seven windows, four in the FUV (far UV) range  (from 1332 to 1358\,\AA\ and from 1389 to 1407\,\AA), and three in the NUV (near UV) range (from 2783 to 2834\,\AA), covering the C\,{\sc ii} doublet at 1335\,\AA, the O\,{\sc i} and C\,{\sc i} lines near 1353\,\AA, the Si\,{\sc iv} doublet near 1400\,\AA, and the Mg\,{\sc ii} h and k lines near 2800\,\AA, as well as two narrow spectral regions in the red wing of  Mg\,{\sc ii} h line. In our data set the sampling step in the direction of the dispersion was 0.025\,\AA, which corresponds to 5.4\,km\,s$^{-1}$ in the FUV band and 2.7\,km\,s$^{-1}$ in the NUV band of IRIS; we note that the resolution of the spectrograph is 0.026\,\AA\ and  0.053\,\AA\ respectively \citep{2014SoPh..289.2733D}. Along the slit, the sampling step was 0.17\arcsec, which is about two times smaller than the effective spatial resolution of IRIS ($\approx0.4$\arcsec). The width of the slit was 0.33\arcsec\ and the exposure time 8 seconds.

Slit-jaw (SJ) images were obtained in the 1400\,\AA\ wavelength band, with a spectral width of 55\,\AA, which includes the Si\,{\sc iv} lines and in the 2796\,\AA\ band (Mg\,{\sc ii} k line) with a width of 4\,\AA. The effective field of view was 114\arcsec$\times$122\arcsec, the sampling step 0.17\arcsec, the cadence 19 seconds and the exposure time 8 seconds. The 1400\,\AA\ images were taken when the slit was at position 1 and the 2796\,\AA\ images when the slit was at position 2; this makes a total of 318 images at each wavelength.

For better visibility of the solar features, the slit was removed from the SJ images, using the temporal and spatial average of its intensity profile; the removal was very efficient, as can be seen in the figures of the following sections. The measured full width at half maximum (FWHM) {of the slit} was 0.9\arcsec\ {in the 2796\,\AA\ SJs and 0.6\arcsec\ in the {1400}\,\AA\ SJs}, larger than the nominal value quoted above; the distance between the two slit positions, measured by cross correlation, was 0.9\arcsec, rather than the nominal 2\arcsec. The cross correlation of the SJ images with the {\it Atmospheric Imaging Assembly} (AIA) 1600\,\AA\ images revealed a slow position drift with a period comparable to the duration of the observing run and an amplitude of 0.9\arcsec\ perpendicular to the slit and 0.3\arcsec\ parallel to the slit; moreover, we found that the slit was tilted by $-0.75$\degr\ with respect to the NS direction. All these corrections were applied to the SJ images used in this work.   

We used level 2 data from the IRIS site, which are properly oriented and corrected for dark current and flat field. Still, further corrections were necessary for pointing and jitter, stray light and absolute calibration, described in Article I and in the previous paragraph. Hot pixels were rather scarce;  when it was deemed necessary, they were removed by comparing each image with the previous and the next one. The position of the limb was determined from the inflection point of the center-to-limb intensity variation at the brightest point of the {IRIS} photospheric spectrum, at 2832.0\,\AA, {and was applied to all spectra and SJ images using the fiducial mark.} The zero of the line-of-sight (LOS) velocity was determined from the average of the \Si\ peak positions over the entire field of view. Our full \Si\ data set is presented in the accompanying movie described in Appendix~\ref{append}.

\section{Overview of a Limb Event}\label{overview}
\begin{figure}[h]
\begin{center}
\includegraphics[height=3cm]{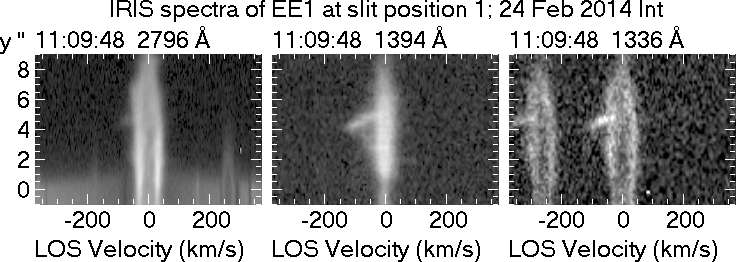}~\includegraphics[width=3.6cm,height=2.37cm]{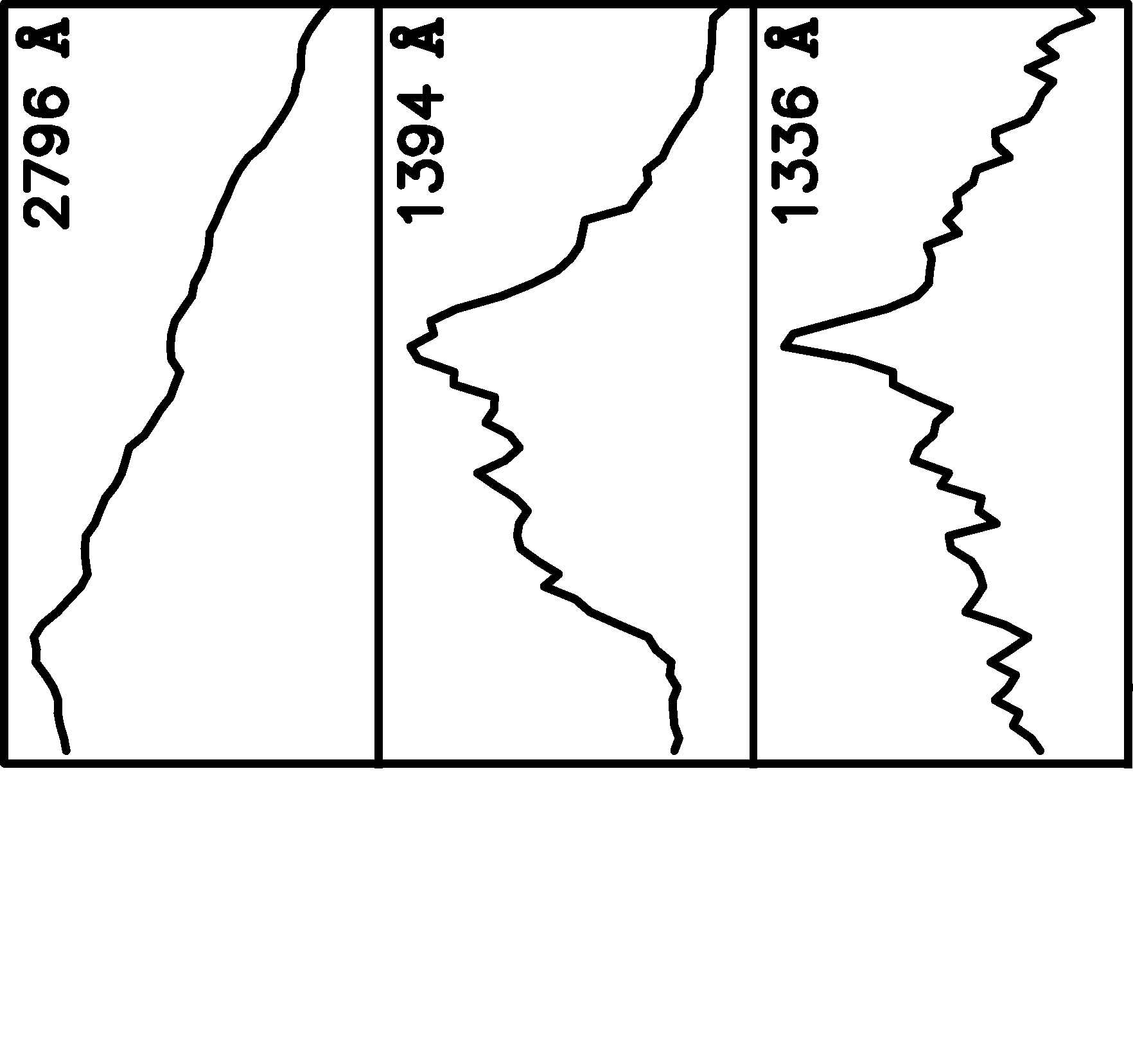}
\end{center}
\caption{Spectra of an explosive event beyond the limb (EE1) in the \Mg\ k line, the \Si\ 1393.79\,\AA\ line, and the \Cb\ doublet. Positions ($y$) are measured with respect to the south limb. {\it At the right}, plots of the integrated intensity {in arbitrary units} as a function of position along the slit, for k, \Si, and \Cb, are given for reference. The position range is the same as in the spectra.}
\label{Spec1}
\end{figure}

In order to illustrate the observational characteristics of EEs, we present in Figure~\ref{Spec1} spectra of an event (EE1) observed 4.5\arcsec\ above the south limb in the \Mg, \Si, and \Cb\ lines. It was best seen in \Si\ and was rather weak in \Mg. It was strongly blue-shifted, with emission detectable beyond $-120$\,\kms, and its spectrum was inclined. {The intensity, integrated over the spectrum, for the same lines, is plotted at the right of the figure; it shows clearly the event in \Si\ and \Cb\ and less so in the \\Mg k line.} We note that this and other EEs had no effect on the lower-temperature lines accessible by IRIS.

\begin{figure}[h]
\begin{center}
\includegraphics[width=.9\textwidth]{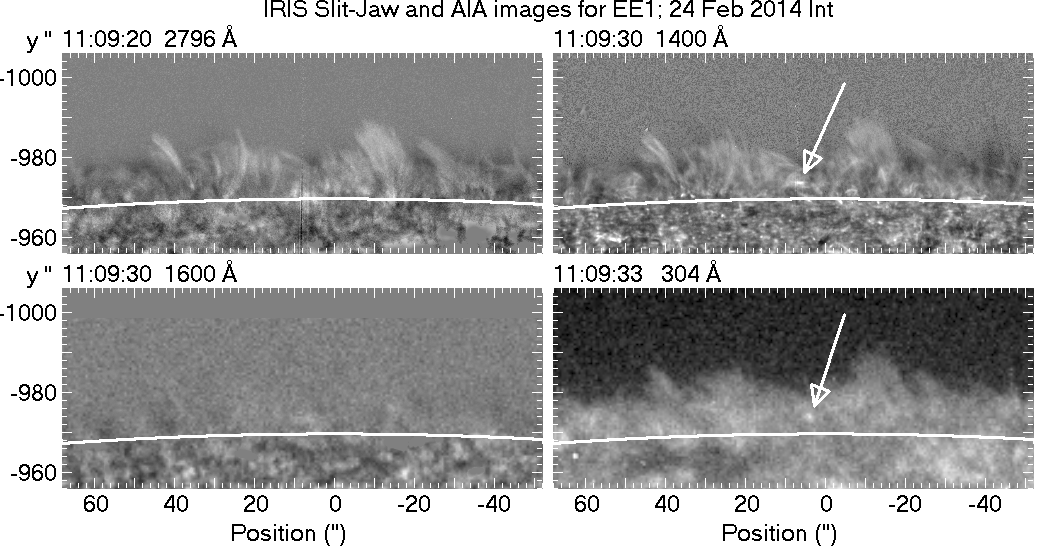}
\end{center}
\caption{Slit-jaw and AIA images during the explosive event of Figure~\ref{Spec1}. {\it Top row\/}: IRIS SJ images in the {2796\,\AA\ } and the {1400\,\AA\ bands}. {\it Bottom row\/}: AIA images in the 1600\,\AA\ and 304\,\AA\ bands. The time-averaged intensity has been subtracted from the 2796, 1400, and 1600\,\AA\ images, to enhance the visibility of the features. Positions are measured from the disk center and solar South is up. The {\it white arc} marks the photospheric limb. The {\it arrows} point to the {explosive} event.}
\label{SJ1}
\end{figure}

The corresponding SJ 2796 and 1400\,\AA\ images are shown in Figure~\ref{SJ1}, together with AIA images in the 1600 and 304\,\AA\ bands. We note that the EE is visible as a weak bright feature in the 1400 and 304\,\AA\ images (arrows), but it is invisible in the 2796\,\AA\ image, as well as in the 1600\,\AA\ image. The lack of emission in the 2796\,\AA\ image probably indicates that the EE is behind the optically thick spicule forest, and this also explains its weak visibility in the \Mg\ spectra.

\section{Detection of Explosive Events and Their Height Distribution}\label{height}
From the example presented {in Figure~\ref{Spec1}} above, it is obvious that the far wings of the \Si\ lines (beyond $\pm50$\,\kms), are the most suitable for the detection of EEs (see also the accompanying movie).  Indeed, the accompanying movie shows a plethora of such events. {Thus, for their detection, we} created images of spectral intensity as a function of time and position {by integrating the intensity over 50\,\kms\ wide spectral windows, centered at $-100$\,\kms\ and 100\,\kms\ from the core of the \Si\ lines;} this velocity range was chosen in order to be at a safe distance from the line center. In order to reduce noise, we averaged over both lines of the \Si\ doublet, at 1402.80 and 1393.79\,\AA.

\begin{figure}[h]
\begin{center}
\includegraphics[width=\textwidth]{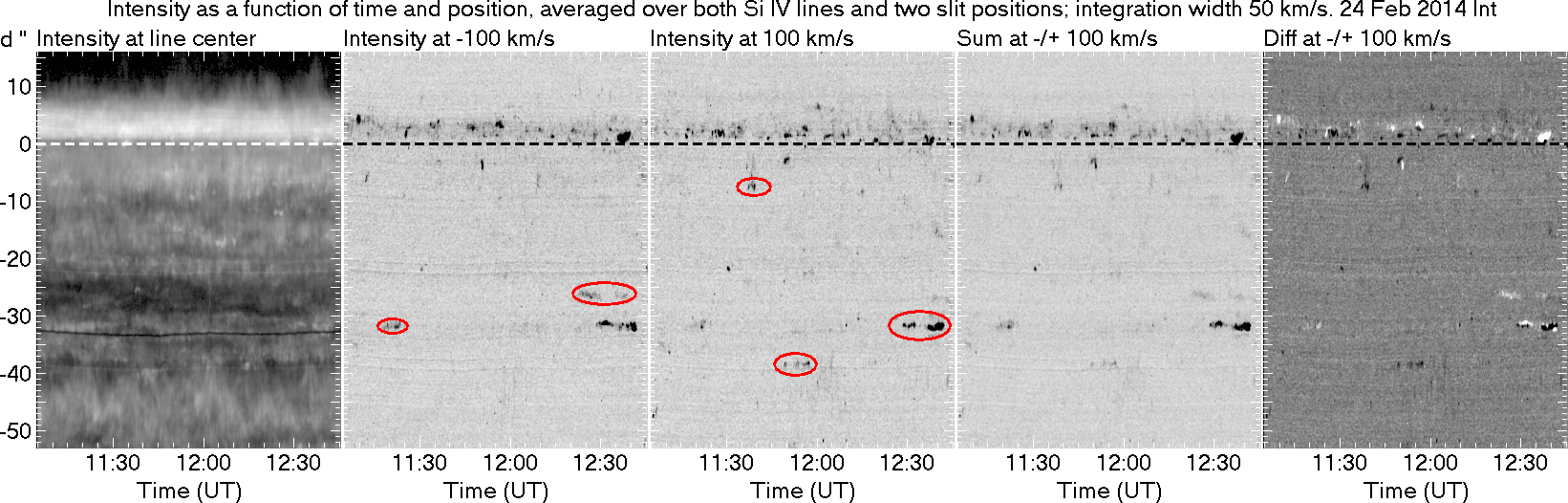}
\end{center}
\caption{Intensity as a function of time and distance from the limb at line center, at $\pm$100\,\kms\ from the center of the \Si\ lines, their sum and their difference. Blue shifted features are bright in the difference image. {\it Red ellipses} mark trains of EEs.}
\label{SiIV100}
\end{figure}

The resulting blue- and red-wing images, averaged over the two lines of the \Si\ doublet and over the spectra taken at the two slit positions, are shown in Figure~\ref{SiIV100}, together with line-center images and images of their sum and difference. We note a significant concentration of EEs beyond the limb going as high as 10\arcsec, as well as at {some} positions on the disk, {e.g. near $-5$\arcsec\ and $-32$\arcsec,} at the location of network structures.

\begin{figure}
\begin{center}
\includegraphics[width=\textwidth]{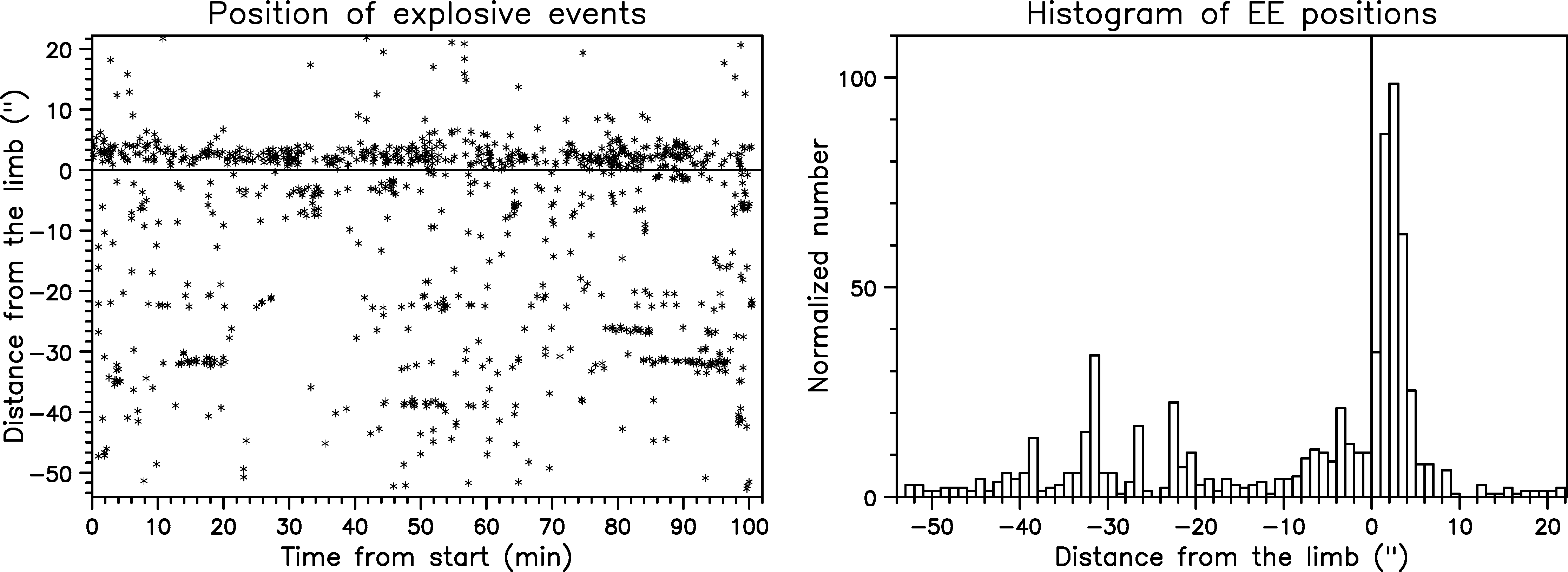}
\end{center}
\caption{Location of explosive events. {\it Left\/}: Position of events detected in the sum of \Si\ images and both slit positions, as a function of time and distance from the limb. {\it Right\/}: Histogram of the distances from the limb.}
\label{PosHist}
\end{figure}

Our next step was to write a simple algorithm for the detection {of the EEs and the measurement of their location} on the images of Figure~\ref{SiIV100}. The {intensity of the detected EEs was} above the 4 $\sigma$ level. Their locations are drawn in the left panel of Figure~\ref{PosHist} for the sum of both \Si\ lines and both slit positions, where 960 EEs are plotted. The right panel of the same figure shows the histogram of their distance from the limb, which verifies our previous statement about the concentration of EEs beyond the limb and at particular locations on the disk.

\begin{figure}[h]
\begin{center}
\includegraphics[width=.6\textwidth]{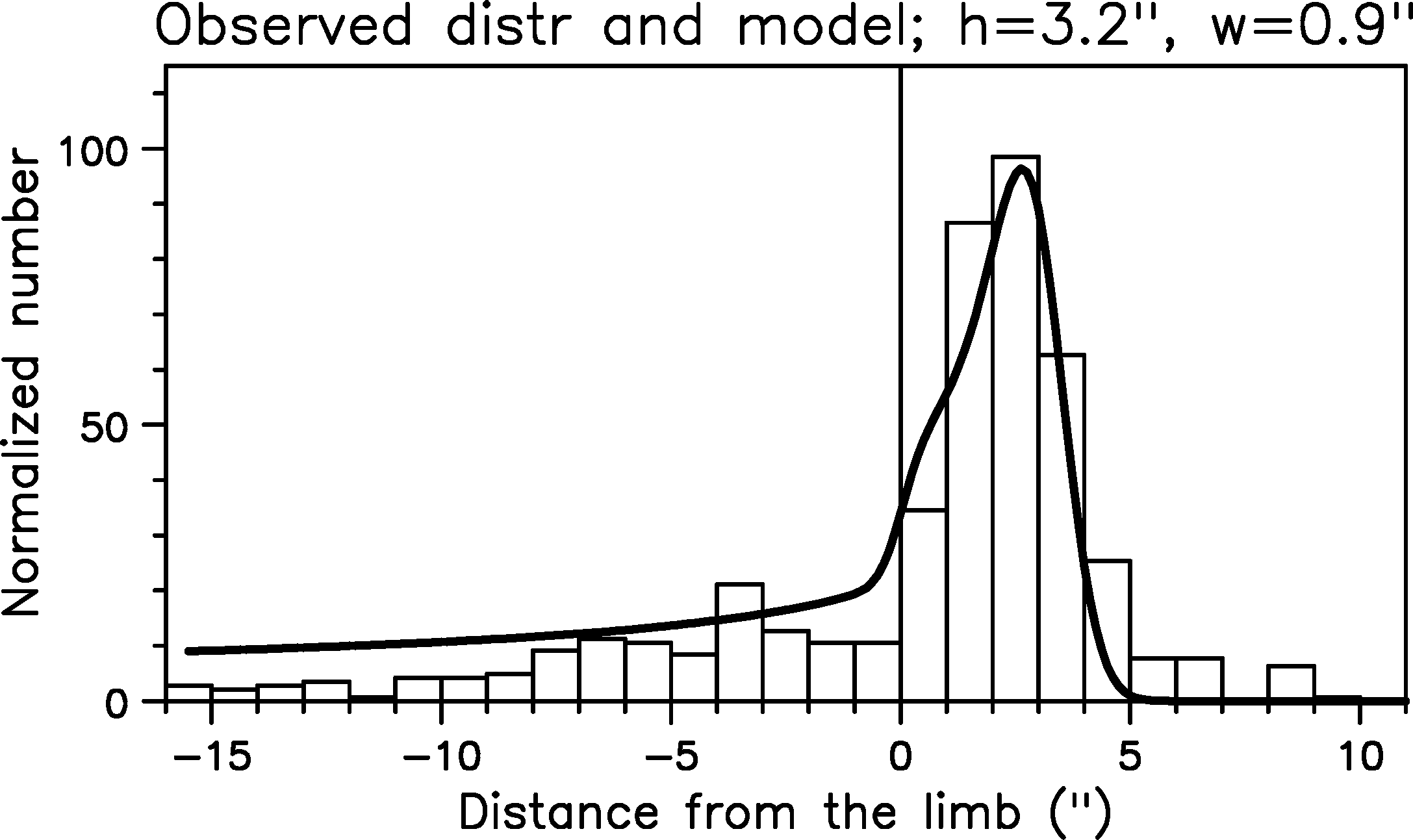}
\end{center}
\caption{Observed EE distribution and model ({\it thick line\/}). {The {\it vertical line} is at the limb.} }
\label{PosMod}
\end{figure}

{As the measured distribution of EE positions reflects both the actual height distribution of the events and projection effects,
we modeled the observed distribution by assuming}  a Gaussian height distribution [$f(z)$] with an average height $h$ and full width at half maximum $w$:
\be
f(z)=f_0\exp[-\ln16(z-h)^2/w^2], \hspace{0.5cm} z>0
\ee
which gives the observed distribution, $F(d)$:
\be
F(d)=\int_{-\infty}^{\ell_0} f(z) \mathrm{d} \ell
\ee
where $d$ is the distance from the limb, projected on the plane of sky, $\ell$ is along the line of sight, and the integration goes to $l_0=\infty$ beyond the limb and stops at the $z=0$ on the disk. 

The best fit is displayed in Figure~\ref{PosMod}, where the model distribution has been smoothed to match the bin size of the observations (1\arcsec).  It shows that EEs occur within a narrow height range ($w=0.9$\arcsec) around $h=3.2$\arcsec. We note that, if our events are of the same nature as those reported by \cite{2014Sci...346C.315P}, the deduced average height is contrary to their suggestion that they form deep in the atmosphere.

\begin{figure}[h]
\begin{center}
\includegraphics[width=\textwidth]{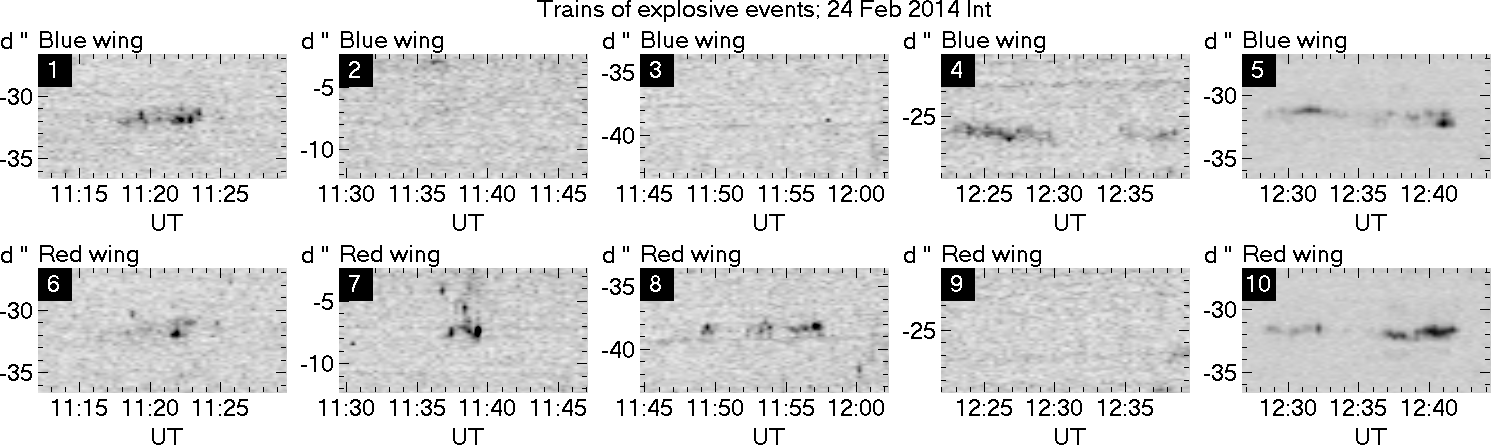}
\end{center}
\caption{Trains of disk explosive events in the blue and red wings of the \Si\ lines.}
\label{trains}
\end{figure}

In addition to the positions, the images of Figure~\ref{SiIV100} provide other important information on explosive events. The wing images show blue/red asymmetries and reveal some instances with consecutive explosive events (trains), marked with red ellipses in Figure~\ref{SiIV100} and shown in detail in Figure~\ref{trains}. The latter, lasting up to about 18\,minutes, consists of many events at about the same position. Repeating EEs have been reported in the past (\citealp{1998ApJ...497L.109C}; \citealp{2004A&A...419.1141N}; \citealp{2006A&A...446..327D}); periodicities were detected in some cases, interpreted in terms of reconnection modulated by {\it p}-mode waves by \cite{2006SoPh..238..313C}. In our case no obvious periodic behavior was found. 
Interestingly, the trains shown  in panels 2 and 7, as well as in panels 3 and 8 appeared in the red wing only, whereas the one in panels 4 and 9 emitted in the blue wing only, thus affirming the asymmetry mentioned above. We will come back to both of these issues below.

\section{Detailed Study of Three Explosive Events}\label{details}

\subsection{The EE Around 11:09:30 UT (EE1)}\label{FirstEvent}
\begin{figure}[h]
\begin{center}
\includegraphics[width=.8\textwidth]{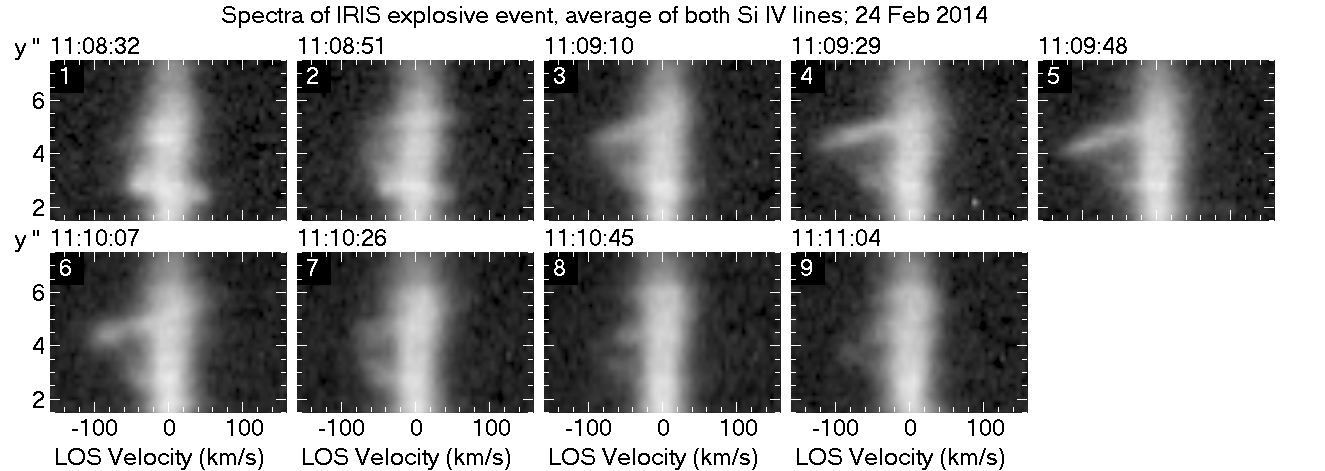}
\end{center}
\caption{A temporal sequence of \Si\ spectra for the first EE at slit position 1{, plotted in a line-of-sight velocity scale}. Both lines of the \Si\ doublet are averaged to reduce noise. The gray scale is the same for all panels.}
\label{SpecSeq}
\end{figure}

\begin{figure}
\begin{center}
\includegraphics[width=\textwidth]{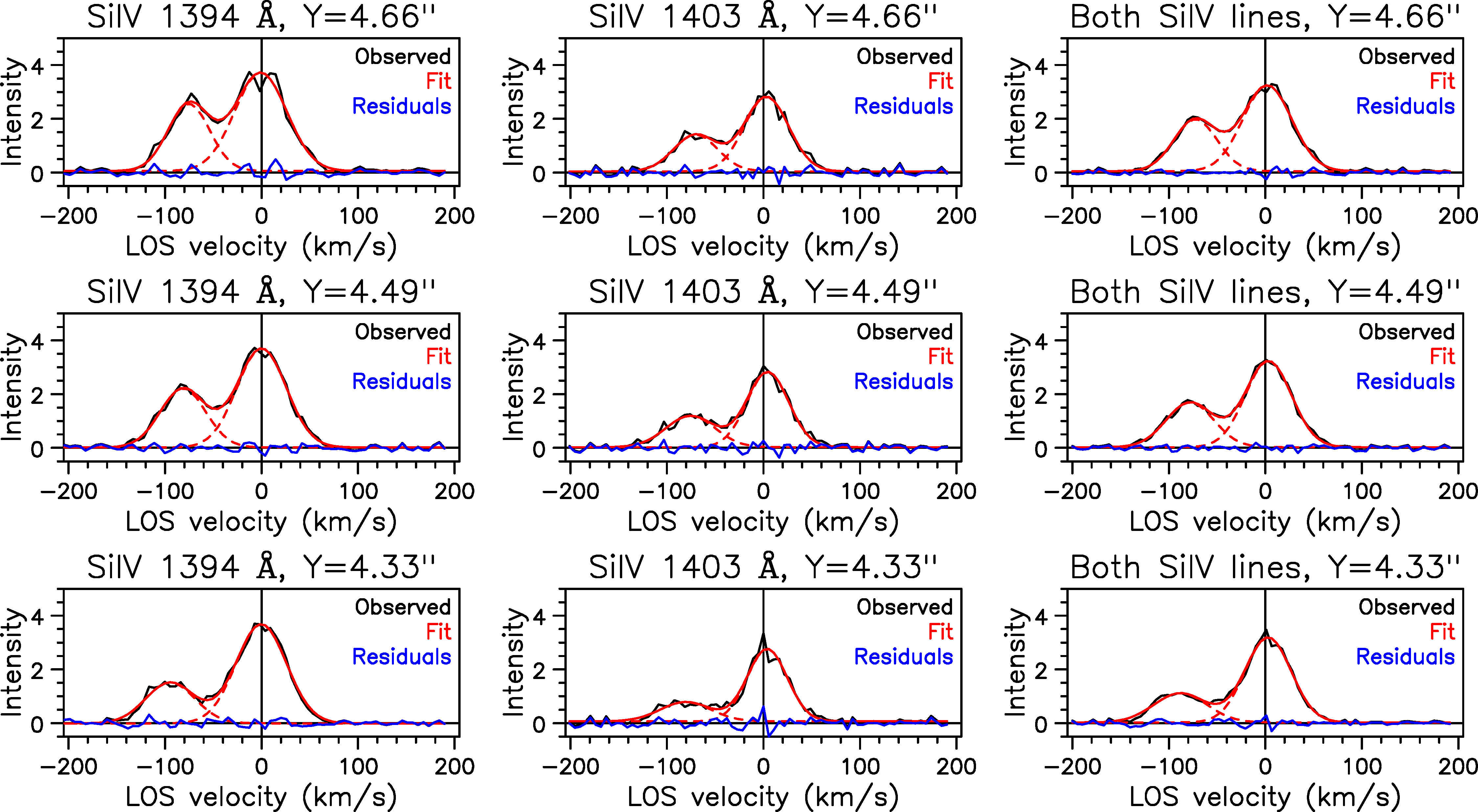}
\end{center}
\caption{Examples of Gaussian fit of \Si\ line profiles at 11:09:29 UT. In this and following figures the intensity is in units of $10^{12}$\,erg\,cm$^{-2}$\,s$^{-1}$\,sr$^{-1}$\,cm$^{-1}$
}
\label{ProfileFits}
\end{figure}

The evolution of the EE presented in Figure~\ref{Spec1} is given in Figure~\ref{SpecSeq} for the sum of the \Si\ doublet spectra. It lasted about three minutes, spanned $\approx1$\arcsec\ around the average height of 4.5\arcsec\ and showed tilted spectra, like the disk event studied by \cite{2017A&A...603A..95A}. The spectra were fitted with the sum of two Gaussian components and representative profiles are shown in Figure~\ref{ProfileFits}, separately for the two lines and for their sum; the original profiles are in black, the fit in red dashed lines, and the residuals in blue. 

\begin{figure}
\begin{center}
\includegraphics[width=.9\textwidth]{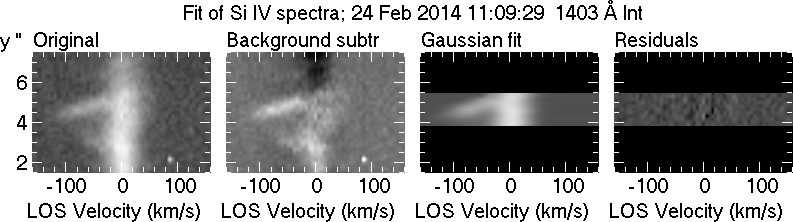}
\end{center}
\caption{
Examples of Gaussian fit of \Si\ line profiles at 11:09:29 UT:
Original spectrum, spectrum with background subtracted, fit, and residuals, all plotted with the same gray scale.}
\label{ProfileFits2}
\end{figure}

We note that the fit is quite good (see also the images in Figure~\ref{ProfileFits2}), with the blue-shifted EE component  well separated from the central component, which comes from the bulk of spicule emission. We further note that some spectra of the stronger \Si\ line, at 1394\,\AA, show a slight central reversal. The temporal average of such profiles was found in Article I to be slightly flat-topped beyond the limb, with a weak minimum at the core. If both lines were optically thin and with the same source function, their intensity ratio should have been equal to two, which is the ratio of their strength. This is not the case (see Figure~\ref{ProfileFits}), and we computed their optical depth [$\tau$] from the intensity ratio; we obtained $\tau\approx1.25$ for the weaker line (1403\,\AA) for the main component and 0.2 to 0.6 for the blue-shifted component; the former is consistent with the value of $\tau=1.5$ reported in Article I for the weaker line at a height of 2 to 4\arcsec\ above the limb.

\begin{figure}[h]
\begin{center}
\includegraphics[width=.7\textwidth]{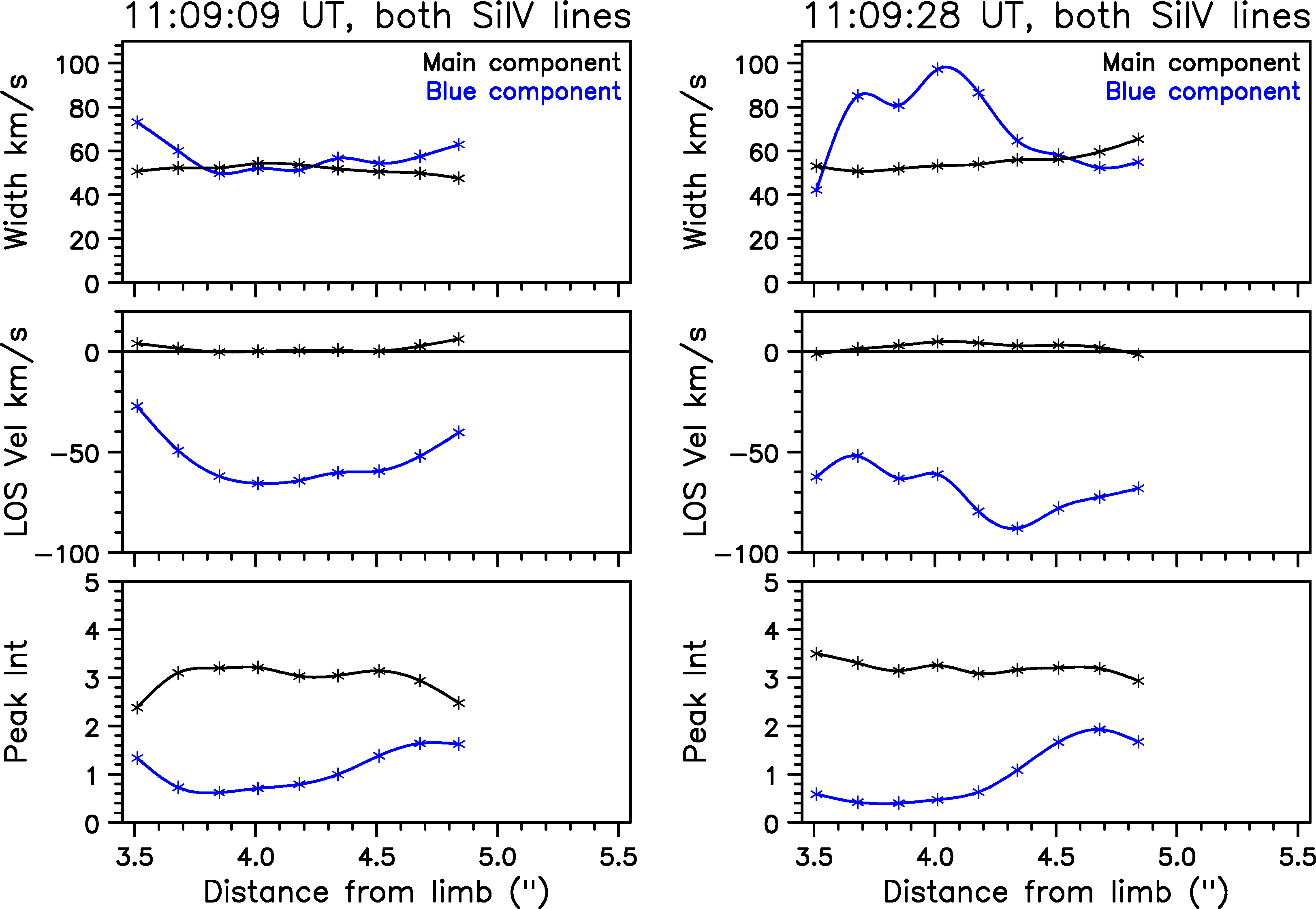}

\smallskip\includegraphics[width=.7\textwidth]{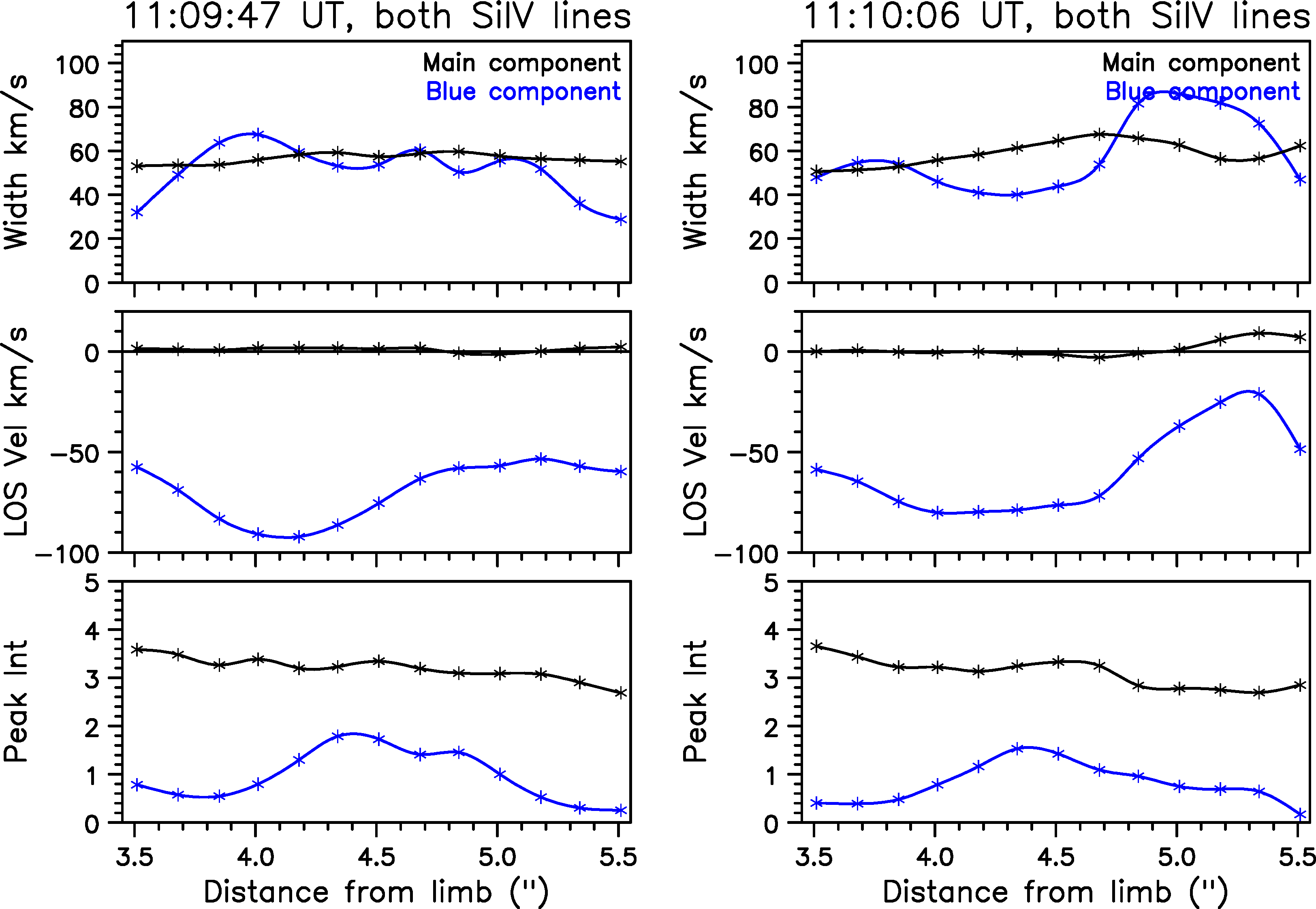}
\end{center}
\caption{Profile parameters as a function of distance from the limb, derived from Gaussian fit, at various times. {\it Blue lines} are for the EE component, {\it black} for the main component. {\it Full lines} are spline curves though the measured values.}
\label{ProfParmT}
\end{figure}

The peak intensity, LOS velocity, and FWHM of the profiles  for some spectra, derived from the Gaussian fit, as a function of distance from the limb, are plotted in Figure~\ref{ProfParmT}. Although the profile of the EE may be partly obscured by the background emission (see panels 3\,-\,6 in Figure~\ref{SpecSeq}), the parameters of the EE are well determined when the two profiles are well separated, as in the cases shown in Figure~\ref{ProfileFits}. However, when the two profiles overlap to a large extent and/or the EE is weak, there can be some cross-talk between the two components. 

We note that the parameters of the background component (black lines) are fairly stable, with a LOS velocity near zero and a slow decrease of peak intensity as a function of distance, reflecting the overall decrease of intensity with height. The two components had comparable spectral FWHM, of the order of 50\kms. The EE component (blue lines) shows a blue shift of up to $\approx-90$\,\kms, with a gradient of about 60\kms\,(arcsec)$^{-1}$. Interpreted in terms of rotation around an axis perpendicular to the slit, this gives a rotation velocity of $\approx30$\,\kms for a 1\arcsec\ wide feature.

From the above analysis we conclude that we are dealing with two features, well separated in most spectra. One is the blue-shifted EE and the other is the unshifted spicular background, which changes little during the event. 

\subsection{The EE Around 12:37:46 UT (EE2)}\label{SecondEvent}
\begin{figure}[h]
\begin{center}
\includegraphics[height=3cm]{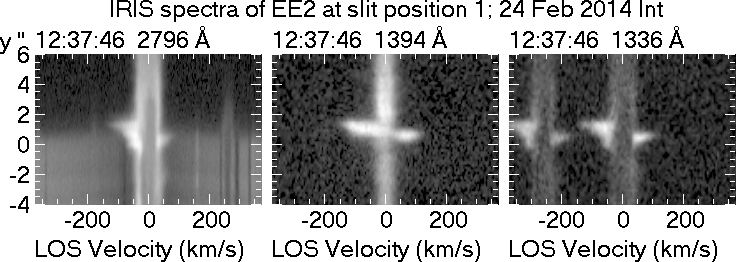}~\includegraphics[width=3.6cm,height=2.37cm]{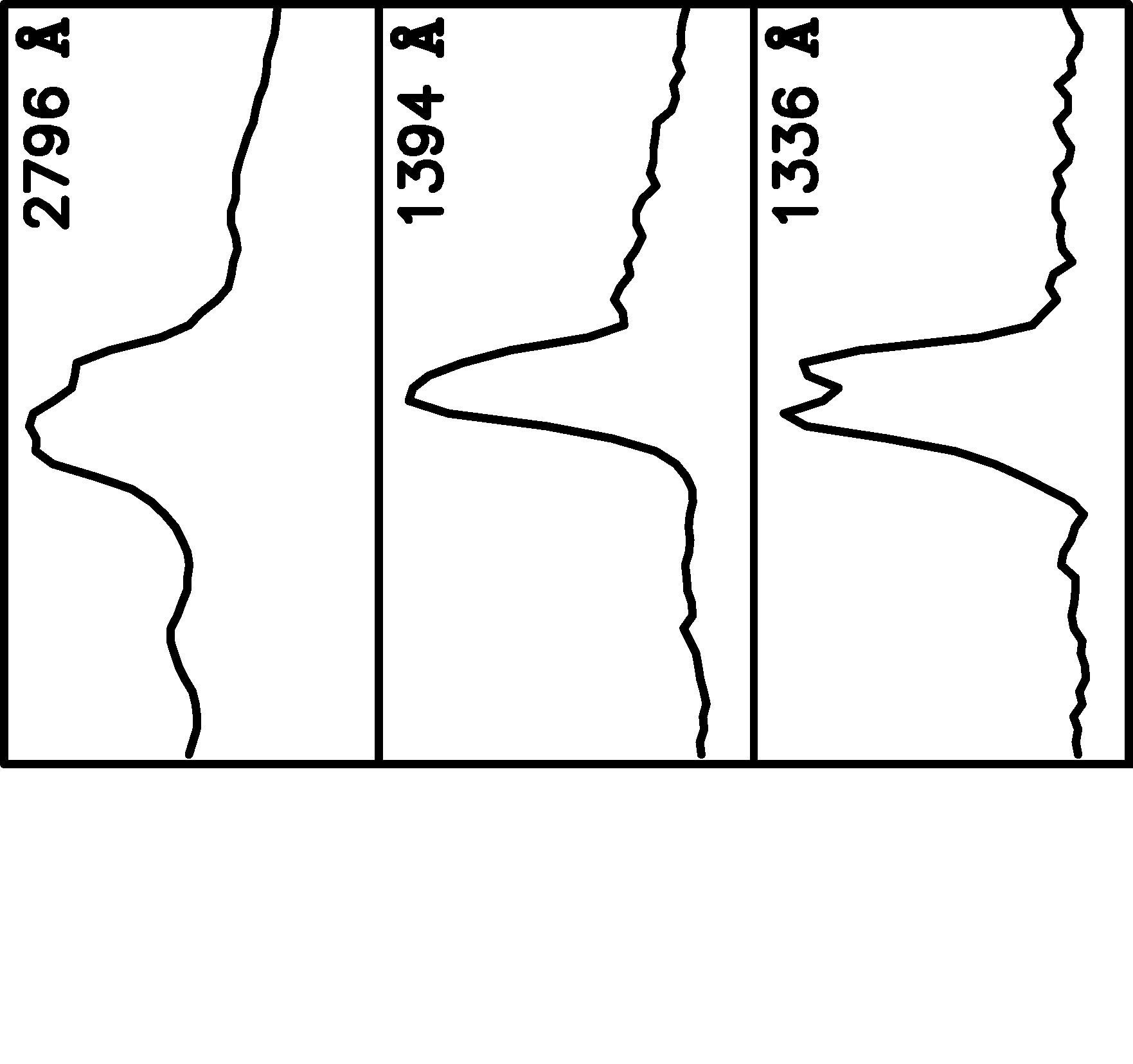}

\smallskip
\includegraphics[width=.9\textwidth]{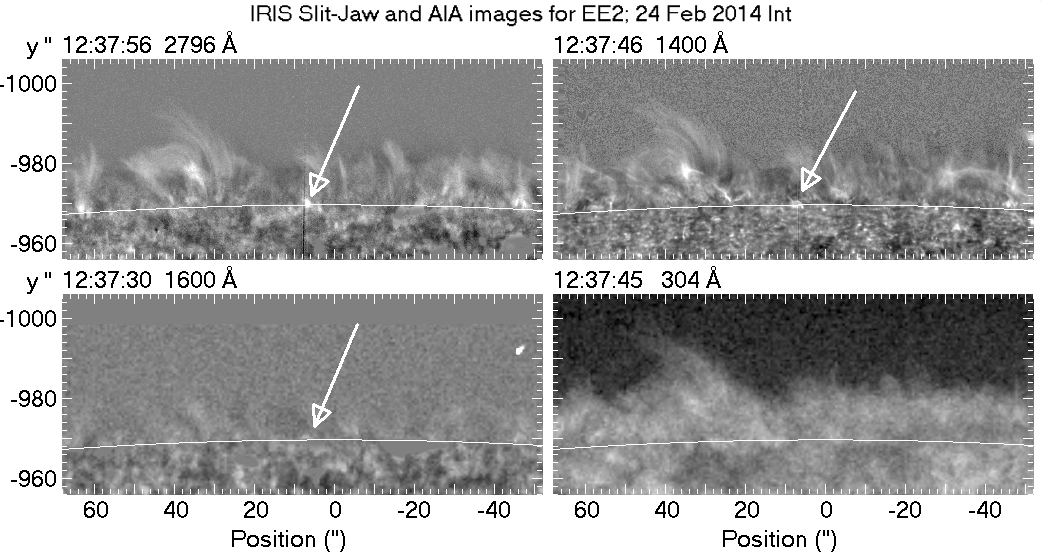}
\end{center}
\caption{{\it Top\/}: Spectra of the explosive event near the limb around 12:37:46 UT at slit position 1.  At the {\it right}, plots of the integrated intensity {in arbitraty units} as a function of position along the slit are given for reference. {The position range is the same as in the spectra.} {\it Bottom\/}: Slit-jaw and AIA images of the same event. The time-averaged intensity has been subtracted from the 2796, 1400, and 1600\,\AA\ images.}
\label{Spec2}
\end{figure}

Spectra of a second EE, which also occurred beyond-the-limb, are presented in the top panel of Figure~\ref{Spec2}, in  the \Mg, \Si, and \Cb\ lines. Its projected height was 1\arcsec, lower than that of the EE discussed in the previous section; moreover, its profile appears more symmetric, with wing emission extending quite far from the line center. The bottom panels of the figure show slit-jaw and AIA images; this time the event is visible in the 2796\,\AA\ and 1400\,\AA\ \Si\ IRIS bands, as well as in the 1600\,\AA\ AIA band, but not in 304\,
\AA.

\begin{figure}
\begin{center}
\includegraphics[width=.8\textwidth]{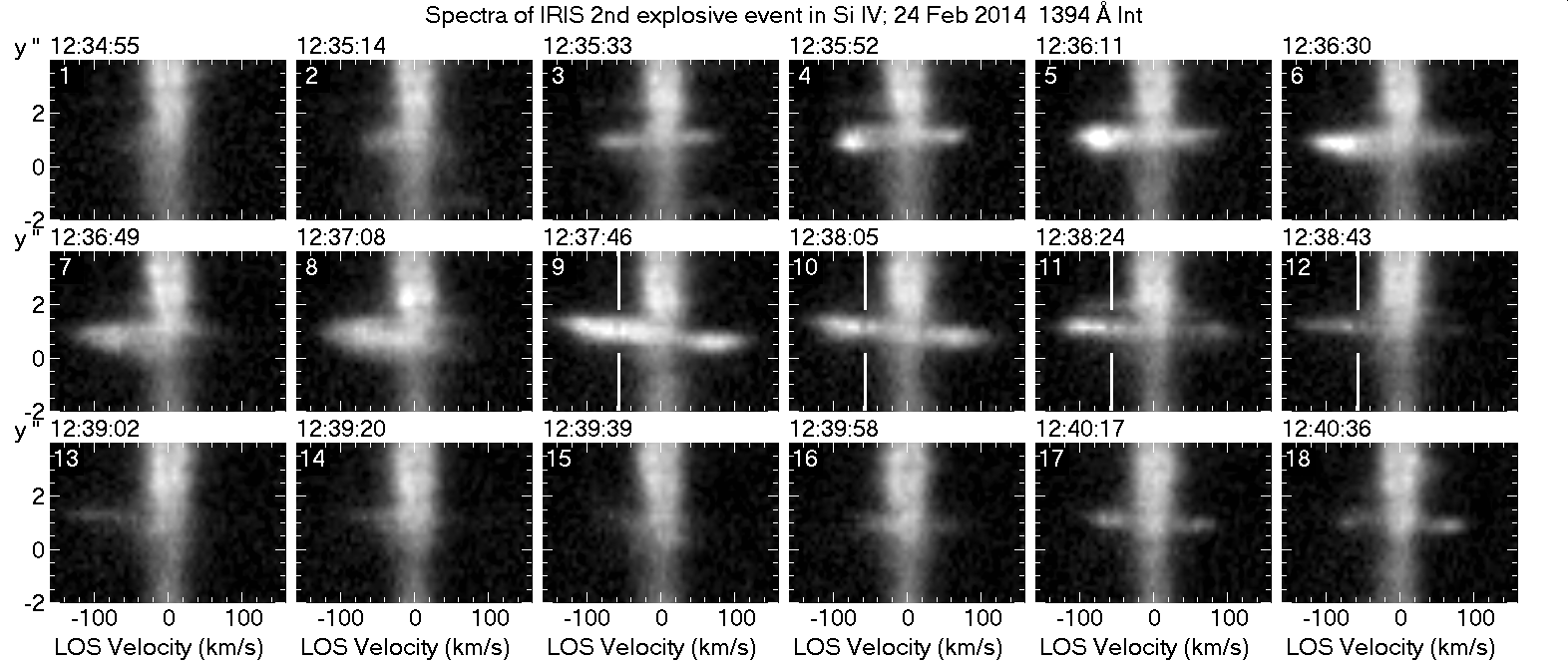}
\end{center}
\caption{A temporal sequence of spectra in the 1394\,\AA\  \Si\ line, for the second explosive event (EE2). The vertical lines in panels 9\,-\,12 mark the absorption feature discussed in the text. The gray scale is the same for all panels.}
\label{SpecSeq2}
\end{figure}

The temporal evolution is shown in the spectral sequence of Figure~\ref{SpecSeq2} for the \Si\ 1394\,\AA\ line. The event started with a strong blue shift (panel 2), then a red-shifted component appeared {with a slight} tilt. Subsequently, both components grew in intensity (panels 4\,-\,5); then, between panels 5 and 6, a slight position shift occurred, probably indicative of a second reconnection event, and gradually a noticeable spectral tilt  {of opposite sense} developed (panels 9-10). While the event was fading (panels 15\,-\,16), another event started (panels 17\,-\,18), practically at the same location.  Thus the evolution was more complex than that of the first EE. We further note that the intensity near the line center did not increase appreciably during the event, which indicates that EE emission in this part of the spectrum was blocked by the foreground plasma; this effect could also explain the absence of emission in the 304\,\AA\ band.

A notable detail is the very weak and narrow (FWHM$\approx9$\,\kms) absorption feature near $\approx-57$\,\kms, marked by the vertical lines in panels 9\,-\,12 of Figure~\ref{SpecSeq2}{; the feature probably exists in panel 5 as well}. As this feature does not appear in the 1403\,\AA\ line of the \Si\ doublet, it is probably not related to the explosive event; as the projected position of the EE is very close to the limb, it could be due to absorption by the foreground plasma in a line formed in the low chromosphere. A possible candidate is the low temperature Fe{\sc\,ii} line at 1393.50\,\AA\ (-62\kms\ from the \Si\ 1393.79\,\AA\ line).

\begin{figure}[h]
\begin{center}
\includegraphics[width=\textwidth]{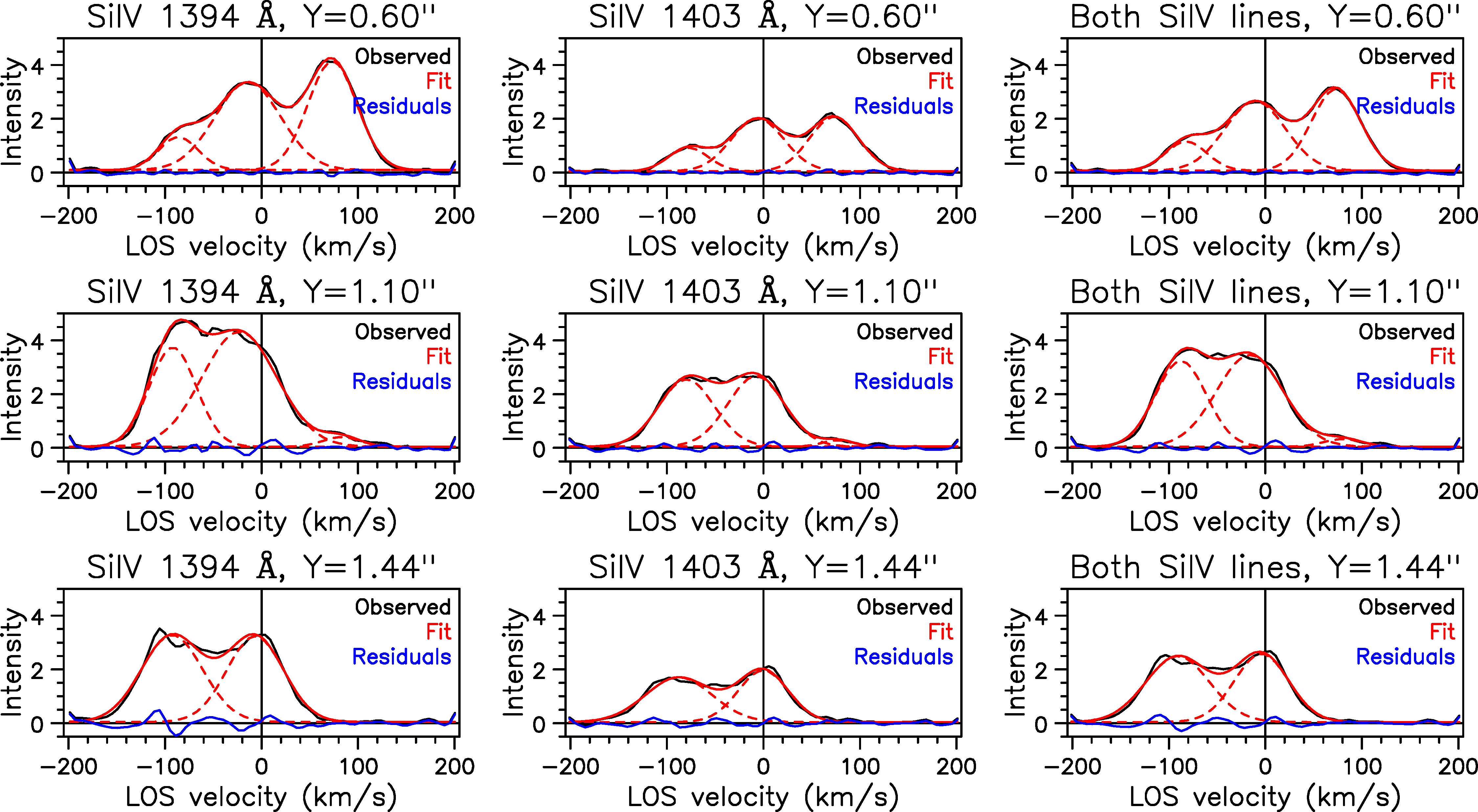}
\end{center}
\caption{Examples of Gaussian fit of \Si\ line profiles at 12:37:45 UT (panel 9 in Figure~\ref{SpecSeq2}) for the second explosive event.}
\label{ProfFit2}
\end{figure}

A detailed line profile analysis of this event is not as straight-forward as it was for the first event. Examples of Gaussian fits of profiles are presented in Figure~\ref{ProfFit2}, for the spectrum at 12:37:45 UT (panel 9 in Figure~\ref{SpecSeq2}). At $y=0.60$\arcsec, three Gaussians were used for the fit, representing the background plasma, the blue, and the red components. Higher up, the red component faded and at $y=1.44$\arcsec\ only two Gaussians were employed. Although at $y=0.60$\arcsec\ and at $y=1.44$\arcsec\ the peaks were fairly well separated, this was not the case at $y=1.14$\arcsec, where the profile appears flat-topped; this effect could induce some cross-talk among the parameters derived from the fit of the three components. Apart from that, the fit appears to be as good as that of the first event (Figure~\ref{ProfileFits}). 

\begin{figure}[h]
\begin{center}
\includegraphics[width=\textwidth]{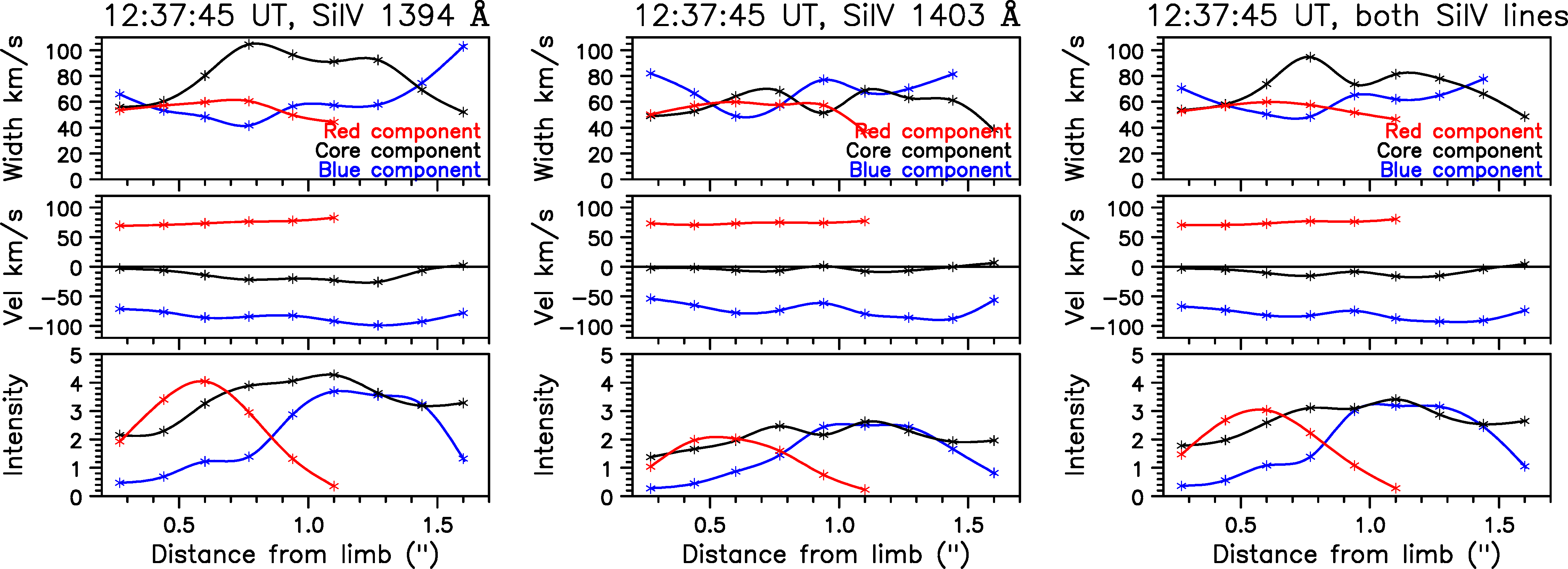}
\end{center}
\caption{Parameters derived from the Gaussian fit of \Si\ line profiles at 12:37:45 UT for the second explosive event, as a function of distance from the limb.}
\label{ProfParm2}
\end{figure}

The intensity, the LOS velocity, and the FWHM of the three components, as a function of position, are plotted in Figure~\ref{ProfParm2}. The intensity and velocity plots show well the spatial and velocity separation of the two shifted components. The red-shifted component displays a small velocity gradient, of about $-20$\kms\,(arcsec)$^{-1}$, a factor of three lower than the gradient of EE1. The gradient of the blue-shifted component, which was located 0.4\arcsec\ higher than the red (panel 9 of Figure~\ref{SpecSeq2}), was smaller and in the opposite sense -- about 12\kms\,(arcsec)$^{-1}$, We note that if the blue- and red-wing components were considered as part of the same line profile, the difference in their location and LOS velocities would give the impression of a spectral tilt of about 200\kms\ (arcsec)$^{-1}$.

An important remark is that all three components have similar line widths. This indicates that the observed large line width is not a real effect, but the result of a mere superposition of three independent structures: the background plasma, the blue, and the red component; 

As in the case of the first event, the optical depth derived from the \Si\ line ratio is smaller than that of the background component, of the order of 1.2 and 2 respectively.

\subsection{An EE on the Disk (EE3)}\label{ThirdEvent}

It will be interesting to check if explosive events on the disk have the same {observational characteristics} as those above the limb, thus in this section  \begin{figure}[h]
\begin{center}
\includegraphics[height=3cm]{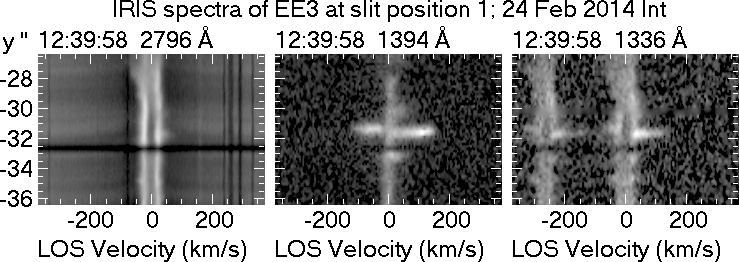}~\includegraphics[width=3.6cm,height=2.37cm]{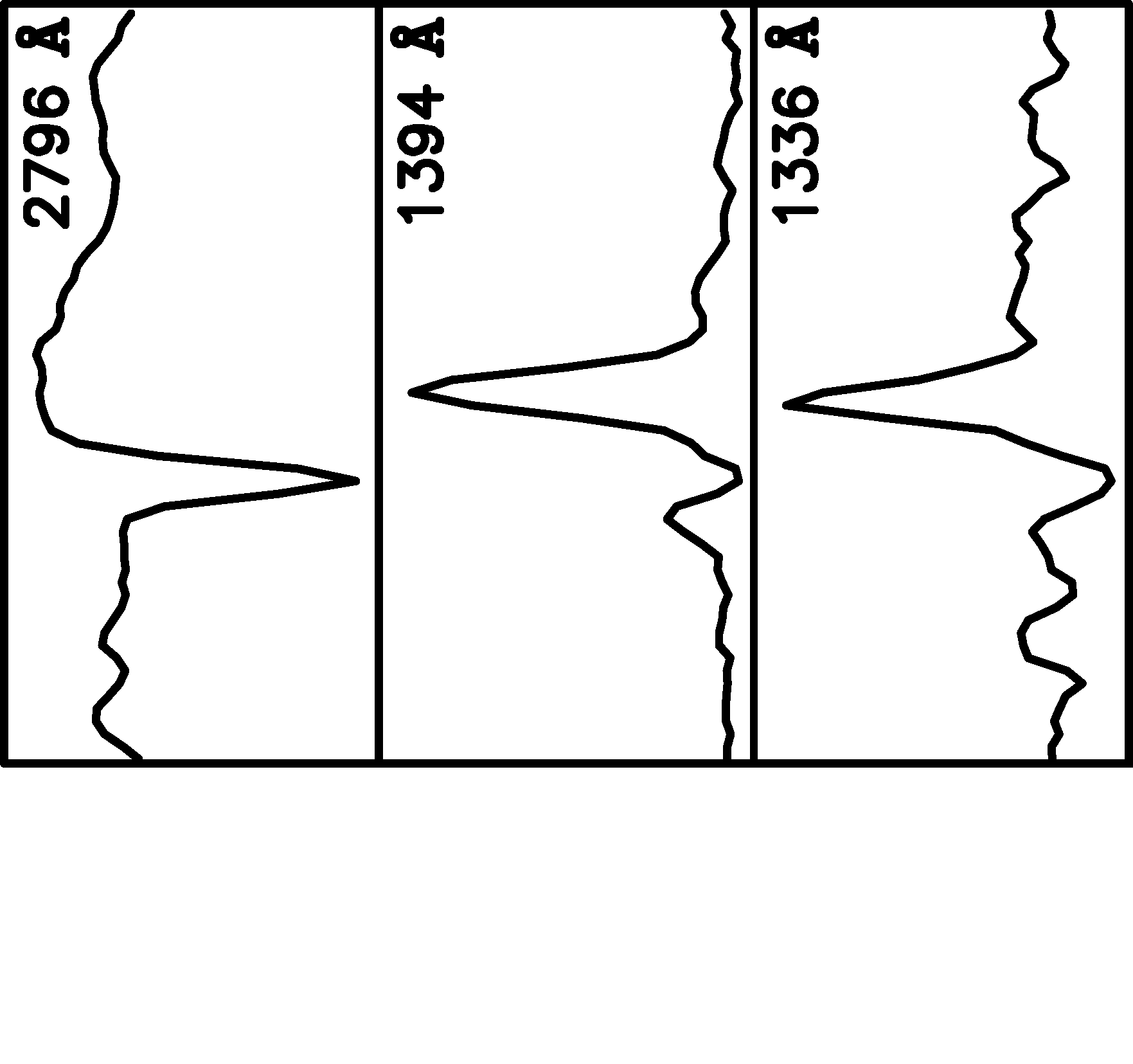}

\smallskip
\includegraphics[width=.9\textwidth]{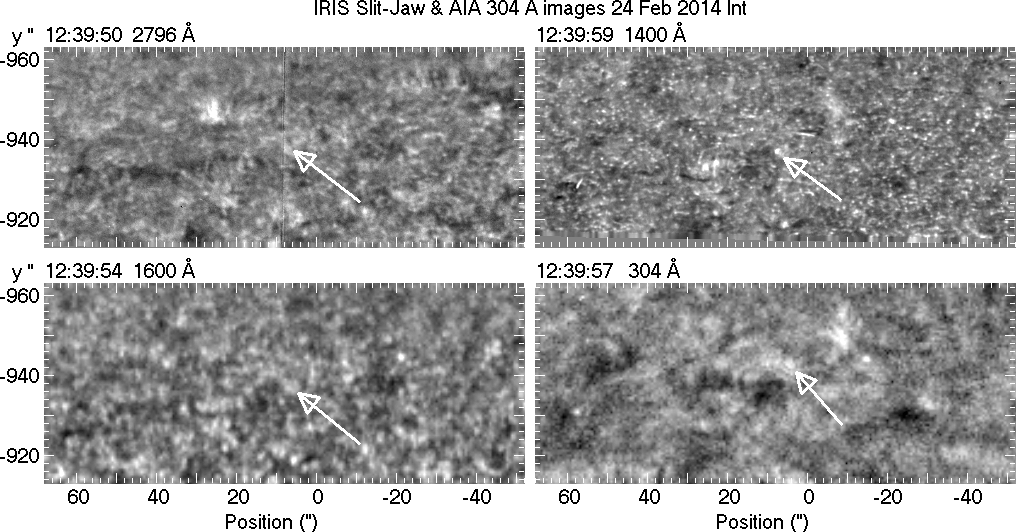}
\end{center}
\caption{An explosive event on the disk. {\it Top\/}: Spectra at 12:39:58 UT at slit position 1. The fiducial mark is at $-32.6$\arcsec. At the {\it right}, plots of the integrated intensity {in arbitraty units} as a function of position along the slit are given for reference. {The position range is the same as in the spectra.}
{\it Bottom\/}: Slit-jaw and AIA images of the same event. The time-averaged intensity has been subtracted from the 2796, 1400, and 1600\,\AA\ images.}
\label{Spec3}
\end{figure}
we will briefly describe the strongest disk event, that occurred 31.4\arcsec\ inside the limb ($\mu=0.25$) and was part of the train shown in the right column of Figure~\ref{trains}. Spectra of this event (EE3) in  
the \Mg\ {(2796\,\AA)}, \Si\ {(1394\,\AA)}, and \Cb\ {(1336\,\AA)} lines at slit position 1 and at the time of its peak are presented in the top panel of Figure~\ref{Spec3}, while SJ and AIA images are shown in the lower panels of the same figure. The spectra show strong emission in \Si, weaker in \Cb, and very weak in \Mg. The event is detectable in the 1400\,\AA\ and 304\,\AA\ images and barely visible in the 2796\,\AA\ and 1600\,\AA\ images; we note that in the 304\,\AA\ image the event is shifted towards the limb, a consequence of the fact that the emission in this spectral band forms higher than in the others \citep{2019SoPh..294..161A}.

\begin{figure}[h]
\begin{center}
\includegraphics[width=.7\textwidth]{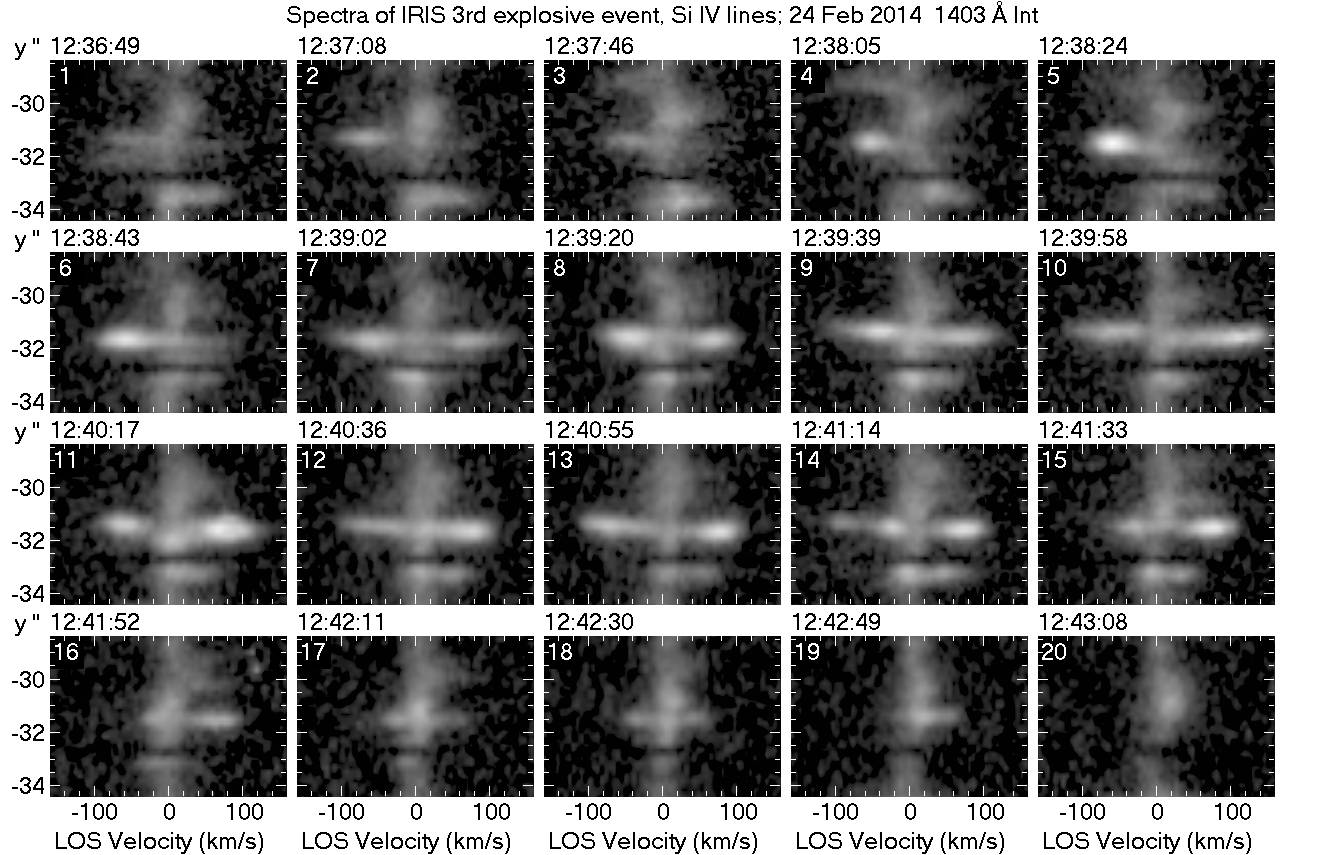}
\end{center}
\caption{A temporal sequence of spectra, averaged over both \Si\ lines, for the disk explosive event at slit position 1. The gray scale is the same for all panels.}
\label{SpecSeq3}
\end{figure}

Although the \Si\ profile appears symmetric in the spectra of Figure~\ref{Spec3}, a closer look at its spectral temporal evolution (Figure~\ref{SpecSeq3}) reveals that this is not the case. As in the case of the second limb event, a blue component appeared first (panel 2), followed by a red component (panel 7), about three minutes later. The blue component decayed first, about two minutes before the red. As for the \Si\ doublet line ratios, they are close  to two, all components being optically thin. Thus the main difference between this disk EE and those beyond the limb described previously is the smaller optical depth of the background component.

\section{Temporal Evolution}\label{TimEvol}
Although the temporal evolution of the explosive events can be followed in the spectral sequences shown in Figures \ref{SpecSeq}, \begin{figure}[!h]
\begin{center}
\includegraphics[height=7.3cm]{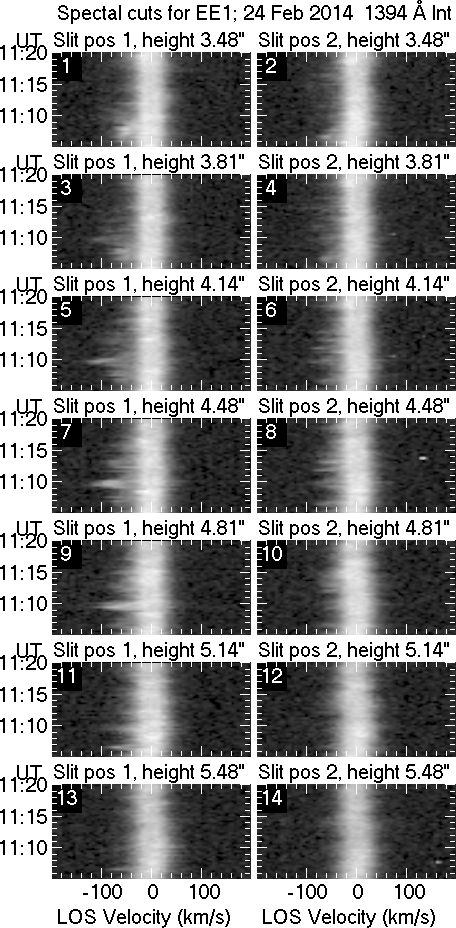}
\includegraphics[height=7.3cm]{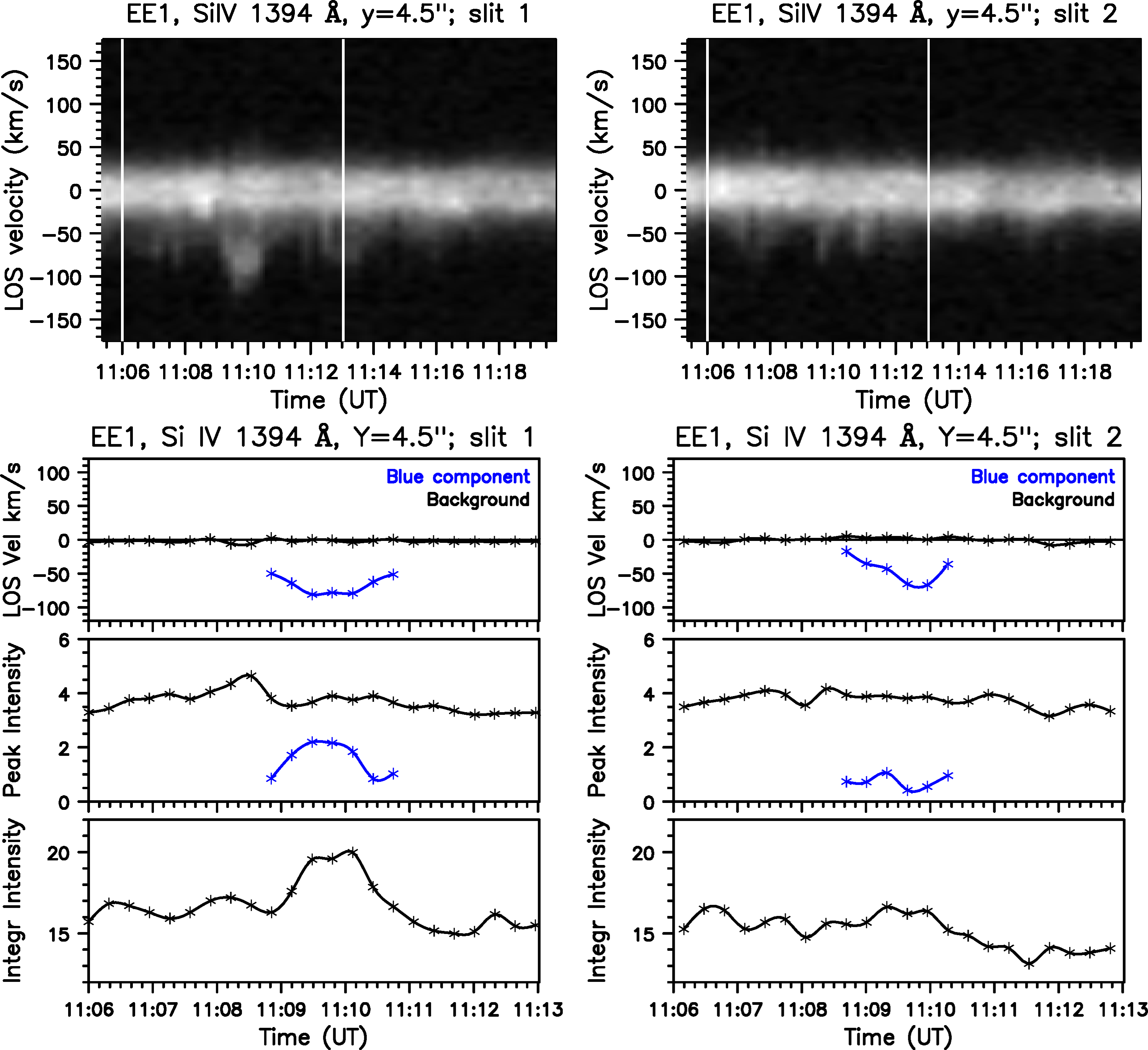}
\end{center}
\caption{Cuts of \Si\ spectra for the first EE at both slit positions. {\it Left\/:} Cuts at various heights above the limb. The gray scale is the same for all panels. {\it Middle and right, top\/:} Expanded and rotated view of the cut at $y=4.5$\arcsec\ at slit positions 1 and 2, respectively. {\it Middle and right, bottom\/:} Integrated intensity, peak intensity, and LOS velocity as a function of time,  at slit positions 1 and 2, respectively. The time segments are 7 minutes long in the plots and 15 minutes long in the cuts. {\it Vertical lines} in the cuts mark the temporal range of the plots.}
\label{Cuts1}
\end{figure}
\ref{SpecSeq2}, and \ref{SpecSeq3}, it is better visible in images of the spectral intensity as a function of LOS velocity and time (spectral cuts), presented in this section for both slit positions. Such images reveal the multiplicity of the events and can be used to measure the acceleration of the plasma.

The first explosive event (left column in Figure~\ref{Cuts1}) was the simplest of the three, with a single blue-shifted component, in addition to the chromospheric (spicular) background. The peak intensity and the LOS velocity derived from the Gaussian fit of the line profiles, for the cut at 4.45\arcsec\ above the limb (panel 7 of Figure~\ref{Cuts1}, left), are plotted as a function of time in the bottom panel of the middle column of the figure, together with the intensity integrated over the entire spectral feature. The top panel of the middle column of the same figure shows the corresponding cut, this time displayed as a function of time and LOS velocity, i.e., transposed with respect to the display in panel 7, and with an expanded time scale.

\begin{figure}[h]
\begin{center}
\includegraphics[height=7.3cm]{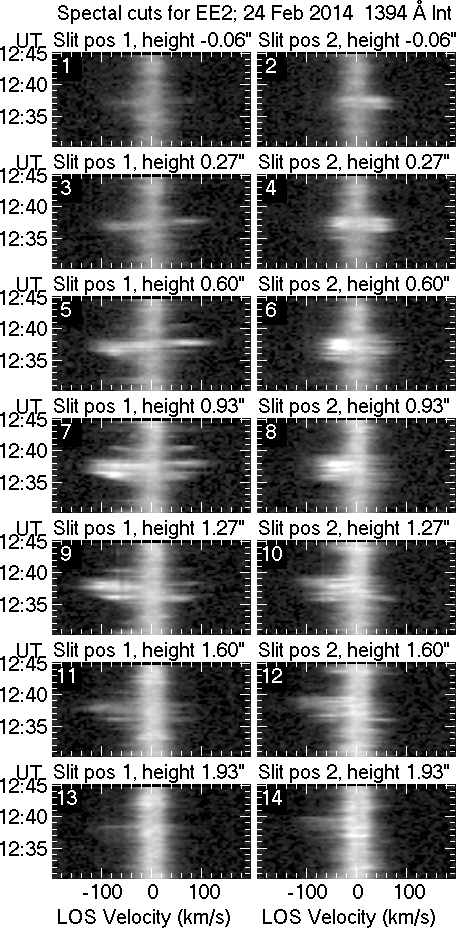}
\includegraphics[height=7.3cm]{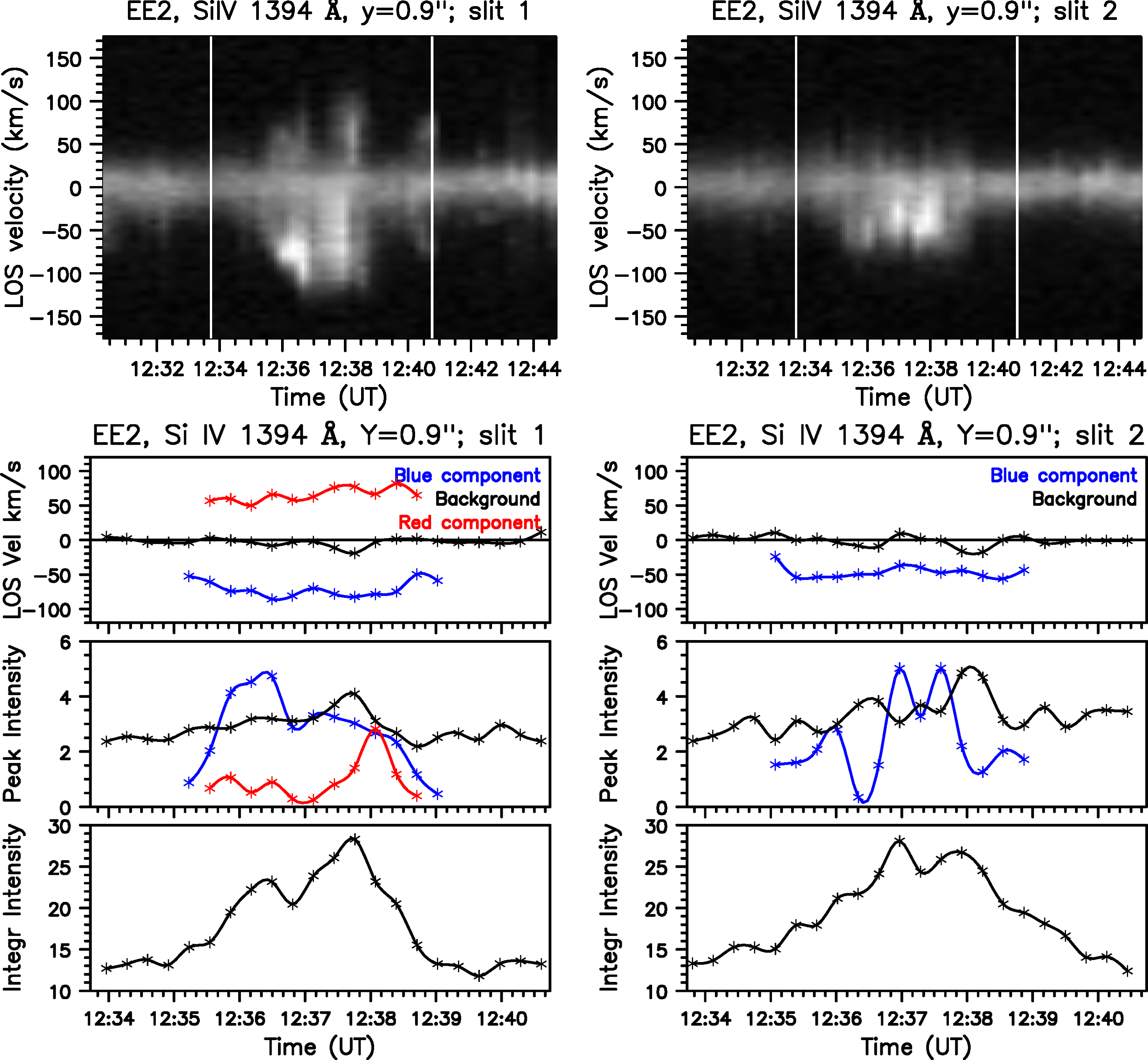}
\end{center}
\caption{Cuts of \Si\ spectra for the second EE at both slit positions. {\it Left\/:} Cuts at various heights above the limb. The gray scale is the same for all panels. {\it Middle and right, top\/:} Expanded and rotated view of the cut at $y=0.9$\arcsec\ at slit positions 1 and 2, respectively. {\it Middle and right, bottom\/:} Integrated intensity, peak intensity, and LOS velocity as a function of time,  at slit positions 1 and 2, respectively. The temporal segments are 7 minutes long in the plots and 15 minutes long in the cuts. {\it Vertical lines} in the cuts mark the temporal range of the plots.}
\label{Cuts2}
\end{figure}

As previously mentioned, the chromospheric component was fairly stable during the event, whereas the EE itself (blue component) had a LOS velocity of $-50$\,\kms\ when first  detected and accelerated to $-80$\,\kms\ two spectra (39 seconds) later, implying an acceleration of $0.80\pm0.04$\,km\,s$^{-2}$. It is interesting that EE1 showed some evidence of acceleration at slit position 2 as well (Figure~\ref{Cuts1}, bottom right), although it was hardly detectable there; we note that the slit position 1 was near the edge of the event (see the figure in the Appendix) and slit position 2 was farther away. The event lasted for about 2 minutes, and during its decay phase the plasma decelerated.

The spectral cuts for the second explosive event (Figure~\ref{Cuts2}) reveal multiple temporal components at slit position 1, two blue shifted and two red shifted, together with the stable chromospheric component. In its rising phase the first blue component showed an accelerating LOS velocity of $0.42\pm0.06$\,km\,s$^{-2}$. At the second position of the slit, EE2 did not have any detectable red component, but did show four blue-shifted components.

\begin{figure}[h]
\begin{center}
\includegraphics[height=7.3cm]{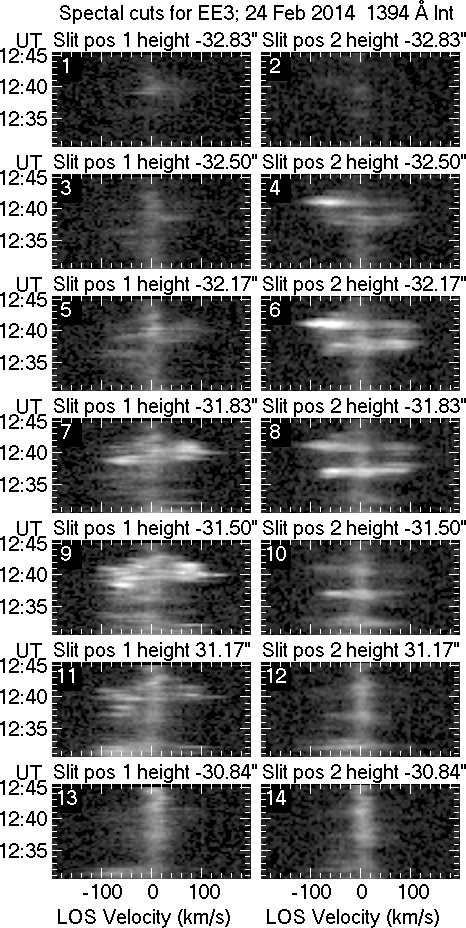}
\includegraphics[height=7.3cm]{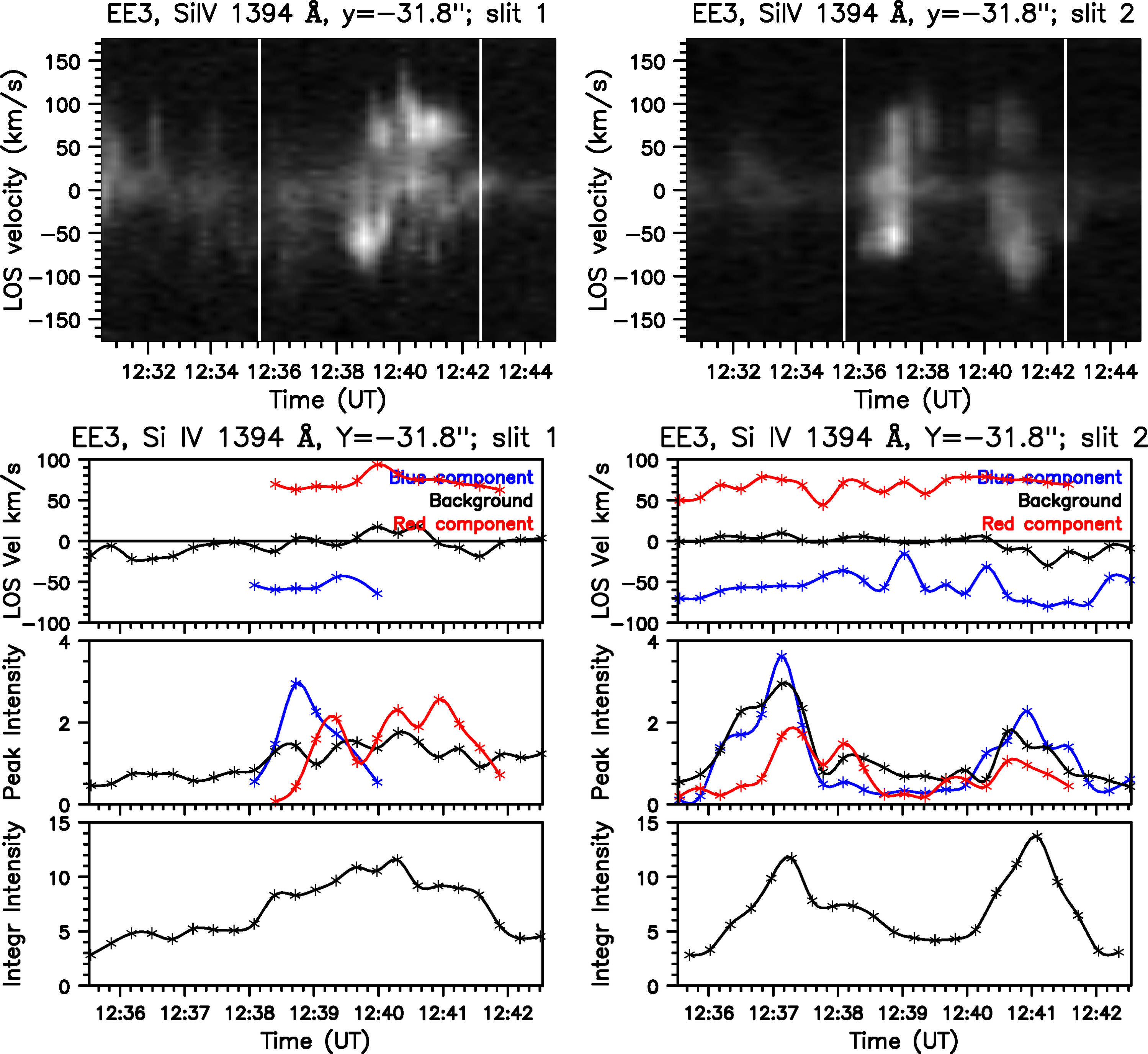}
\end{center}
\caption{Cuts of \Si\ spectra for the third EE at both slit positions. {\it Left\/:} Cuts at various heights above the limb. The gray scale is the same for all panels. {\it Middle and right, top\/:} Expanded and rotated view of the cut at $y=-31.8$\arcsec\ at slit positions 1 and 2, respectively. {\it Middle and right, bottom\/:} Integrated intensity, peak intensity, and LOS velocity as a function of time,  at slit positions 1 and 2, respectively. The temporal segments are 7 minutes long in the plots and 15 minutes long in the cuts. {\it Vertical lines} in the cuts mark the temporal range of the plots.}
\label{Cuts3}
\end{figure}
 
EE3 (Figure~\ref{Cuts3}) was the most complex event of all three, with multiple blue- and red-shifted components at slit position 1. Here the chromospheric background was weak, as we are on the disk, and more variable than in the case of the limb events. It is thus possible that some of the intensity variations near the line center were associated with impulsive episodes; such episodes might be obscured by the background chromospheric emission in limb events. The appearance of the event at slit position 2 was very different, with the components seen at slit position 1 very weak and a new component appearing around 12:37 UT (upper panels, middle, and right columns, in Figure~\ref{Cuts3}). This particular component showed emission around the line center, as well as in the wings; still, the line profiles of this feature are well fitted with three Gaussian sources, much better than with a single broad-band source. 

The complexity of the event does not allow any reliable measurement of acceleration. Still,  in some cases, wing emission with a  LOS velocity of about 60\,\kms\ appeared suddenly, between two consecutive spectra; this puts a lower limit of about 3\,km\,s$^{-2}$ to the acceleration.

\section{{Motions on the Plane of Sky}}
{Concise information about the structure and evolution of the EEs on the plane of sky can be obtained from cuts of the intensity as a function of time and position. Such cuts through the center of the three\begin{figure}[h]
\begin{center}
\includegraphics[width=.6\textwidth]{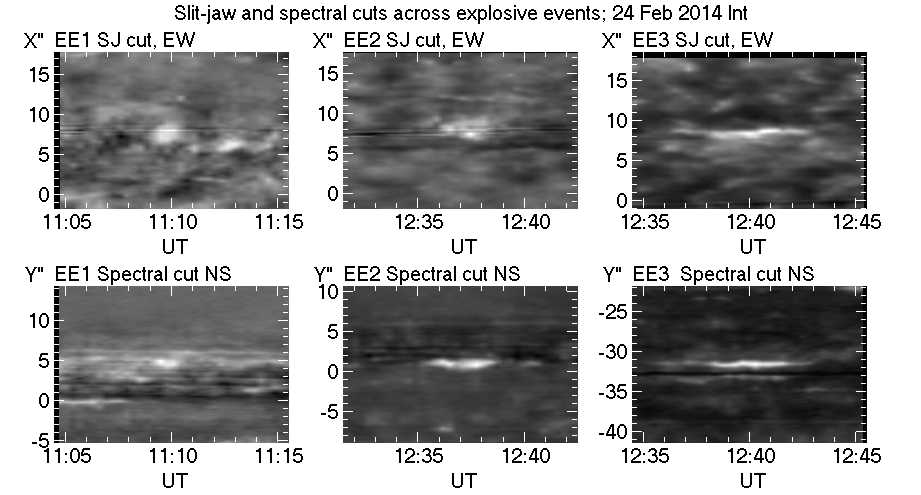}
\end{center}
\caption{{Structure and time evolution of the three EEs studied in detail. The {\it top row} shows images of intensity as a function of time and position along EW SJ image cuts (perpendicular to the slit) at 1400\,\AA, through the center of the events. The slit has been removed, as described in Section~\ref{obs}. The {\it bottom} row shows the spectral intensity along the slit (NS direction) integrated over wavelength. $X$ is measured from the central meridian,  $Y$ is measured from the S limb.}}
\label{motions}
\end{figure} EEs studied here, in the EW (perpendicular to the slit) and NS (along the slit) directions, are shown in Figure~\ref{motions}; SJ images were used for the EW cuts, spectra for the NS cuts.}
 
{A first thing to notice is that there is no indication of structures reminiscent of jets, which were reported to be associated with EEs in some previous works (e.g. \citealp{2019ApJ...873...79C}). A second remark is that EEs are practically stationary during their lifetime. Their velocity on the plane of sky was of the order of 5\kms, much smaller than the LOS velocity; this shows that the motions detected in the spectra were well collimated with the line of sight.}

\section{Summary and Conclusions}\label{summary}
Our investigation of explosive events near and beyond the limb from IRIS data focused on two issues: the measurement of their height above the photosphere and the detailed analysis of  \Si\ line profiles of two limb and one disk event. We used the same data set as in Article I, with some additional corrections  for SJ images: we removed the slit and measured its width on the images (0.9\arcsec\ for 2795\,\AA\ and 0.6\arcsec for 1400\,\AA), the distance between the two slit positions (0.9\arcsec), the image drift along and perpendicular to the slit and its orientation with respect to the NS direction ($-0.75$\degr). Although our data sample a very narrow part of the solar disk, the length of the region along the slit is more than 200\,Mm, covering both network and internetwork regions; beyond the limb, the line of sight crosses about 140\,Mm at a projected distance of 5\arcsec, assuming an average chromospheric height of 10\arcsec; therefore, our results are representative of the quiet Sun.

As reported in previous works, our explosive events are characterized by extended emission in the wings of the \Si, \Cb, and \Mg\ lines, well beyond the average line width. They are best seen in the \Si\ doublet, which apparently reflects the fact that the formation height and temperature of the EEs match better those of these particular lines; moreover, the absence of strong central reversal facilitates the interpretation of their line profiles. 

EEs are easily identified in images of the intensity as a function of time and position, in the blue and red wings of the spectral lines. We developed an algorithm for their detection in such images and measured their distance from the continuum limb (at 2832.0\,\AA). Their distribution shows a strong peak just above the limb and some peaks on the disk, associated with chromospheric network structures; assuming a Gaussian distribution of their heights, we found  that they form in a narrow (0.9\arcsec) height range, 3.2\arcsec\ above the photospheric limb. {If our events are of the same nature as those reported by \cite{2014Sci...346C.315P}, this result goes against their suggestion that they form deep in the atmosphere.}

In the detailed analysis of two limb (EE1 and EE2) and one disk event (EE3), we used both spectra and images from IRIS and AIA. The events were weak in the 1400\,\AA\ SJs and even more so in the 2796\,\AA\ SJs, and would have gone unnoticed, were they not detected in the spectra; they were hardly detectable in AIA 1600\,\AA\ and 304\,\AA\ images and invisible in higher temperature AIA bands. Along the slit, the EEs extended over 1 to 1.5\arcsec, {reflecting their vertical extent}; their extent perpendicular to the slit, measured in SJ images, was 1 to 2\arcsec, {reflecting their horizontal size}.

All spectral line profiles were well fitted with Gaussian components. {This is in agreement with the profiles published by \cite{2014Sci...346C.315P} and some profiles in the \cite{2022A&A...657A.132K} collection and in disagreement with profiles published by \cite{2017A&A...603A..95A}, \cite{2019ApJ...873...79C} and some other profiles in the \cite{2022A&A...657A.132K} collection, which were irregular or triangular in shape. We further note that the \Si\ line profile of our disk event is morphologically similar to the IRIS Bomb 1 profile of \cite{2014Sci...346C.315P}. The question of the origin of the profiles of different form warrants further investigation; we note, however, that EEs in our quiet-Sun data set are not as strong nor as wide as some active region events reported in the past (e.g. \citealp{2014Sci...346C.315P};  \citealp{2017A&A...603A..95A})} 

An important finding is that what at first sight appears as broad-band emission is actually a superposition of 2\,-\,3 narrow-band components, with well-separated, though partly overlapping, line profiles. In all cases studied, the background (line center) component coexisted with blue-shifted  (EE1), or blue and red-shifted components; the wing components peaked around 50\,-\,90\,\kms\ and had widths comparable to that of the line-center component. The central component did not show any significant spatial or temporal variation in the limb events, as well as in the disk event EE3 at slit position 1, whereas it followed the temporal variations of the wing components in EE3 at slit position 2; in the case of the limb events, variations of the central component could be obscured by the spicule forest in front of the events (see discussion on optical depth below). 

A very weak narrow-band absorption feature was detected in the spectra of EE2 near $\approx-57$\,\kms, which might be due to absorption by the foreground plasma in the low chromosphere. However, we did not find any strong absorption features in the wing of the \Si\ lines, as reported in previous studies.

An interesting result is that the LOS velocity of the blue-wing component of EE1 decreased with height, with a gradient of about 60\kms\,(arcsec)$^{-1}$, giving a tilt to the spectra; a lower gradient  of about $-24$\kms\,(arcsec)$^{-1}$ was measured in the red component of EE2. These values are above the range of $\pm15$\kms\,(arcsec)$^{-1}$ measured by \cite{1973SoPh...32..345A} for spicules with a slit parallel to the limb. Such gradients could be interpreted in terms of rotation around an axis perpendicular to the slit, in which case the rotation velocity for EE1 would be about 30\,\kms\ for a 1\arcsec\ wide feature; this is about one-third to half of its LOS velocity. In turn, the rotation might be an indication of untwisting of a twisted magnetic-flux tube. 

We measured the optical depth from the intensity ratio of the \Si\ doublet, at 1394\,\AA\ and 1403\,\AA, assuming the same value for the source function of both lines and taking into account that the former is twice as strong as the latter. For limb events and for the central (background) component we obtained $\tau\approx1.25$ for the weaker line (1403\,\AA), consistent with $\tau=1.5$ reported in Article I for this line at a height of 2 to 4\arcsec\ above the limb. On the disk (EE3), the central component was optically thin, with the line ratio close to 2, again consistent with Article I. The wing components had small, but measurable, optical depth, in the range of 0.2 to 1.2; they were thus less opaque than the central component beyond the limb and more opaque than the central component on the disk.

The analysis of the spectral temporal sequence showed that the components of explosive events are discrete, both in space and in time, with EE2 and EE3 having multiple temporal components, with a total duration of about 5 and 7 minutes, respectively, whereas individual episodes and EE1 lasted for about 2\,min. More trains of repeating EEs, lasting up to 18\,min, were detected in position--time images in the line wings (c.f. \citealp{1998ApJ...497L.109C}). In two cases we were able to measure the acceleration of the wing components and obtained values of $0.80\pm0.04$\,km\,s$^{-2}$ and  $0.42\pm0.06$\,km\,s$^{-2}$; in other cases, wing emission with a  LOS velocity of about 60\,\kms\ appeared suddenly, between two consecutive spectra; with a cadence of 19\,s, this puts a lower limit of about 3\,km\,s$^{-2}$ on the acceleration. 

The overall picture that comes out of our analysis is that in the course of EEs, which occur in network boundaries and at an average height of 3.2\arcsec, material is expelled towards and/or away from the observer in discrete episodes in time and in space  ({\it c.f.} \citealp{2019ApJ...873...79C}; \citealp{2022SoPh..297...76T}). The expelled plasma accelerates quickly, reaching  LOS velocities up to 90\kms, sometimes showing rotating motions of up to 30\kms; in EEs with projected position close to the limb, the foreground plasma in the low chromosphere may produce narrow-band absorption features. Interestingly, no motions indicative of jets were detected in SJ or AIA images. {On the plane of sky, the EEs were practically stationary; their velocity was much less than the LOS velocity, showing that their motion was practically along the line of sight.} 

Our study showed that the mass motion must have a significant component parallel to the solar surface {(along the line-of-sight)}, in addition to any vertical component, as we are at, or very close to the limb. Hence reconnection, a prime candidate for the origin of explosive events (see the review of \citealp{2018SSRv..214..120Y} for a discussion of possible scenarios) must have occurred in highly inclined flux tubes, e.g. in the middle or near the top of small loops. Multiplicity in time is probably due to discrete reconnection events in the same or neighboring loops. Profiles with peaks in one wing only can be attributed to asymmetric magnetic field geometry; indeed, if a gradient along the field direction is present in a bidirectional flow, the velocity of the plasma and the intensity of the emission will be stronger at the side of stronger magnetic field. 

We further note that, for limb events, the plasma must be accelerated to a minimum velocity, $v_{\mathrm{min}}$ in order that its emission avoids obscuration by the spicule forest. For a constant acceleration [$a$] this will take a time $t=v_{\mathrm{min}}/a$, in which a distance $s=v_{\mathrm{min}}^2\cos\theta/2a$ will be traversed on the plane of sky, $\theta$ being the inclination of the flow direction with respect to the vertical. This effect will not produce a time difference in the detection of the wing emission, but will result to a position shift of $2s$, giving the impression of two discrete events rather than one. With $v_{\mathrm{min}}=50$\,\kms, $a=3$\,km\,s$^{-2}$ (see above), and $\theta=45$\degr, we obtain a position shift of 1.2\arcsec, well above the IRIS resolution. This effect may explain the position shift between the components of EE2  (panel 9 of Figure~\ref{SpecSeq2}), which could both originate from the same reconnection region.

This work has revealed new properties of explosive events in the quiet Sun, in particular their height, the existence of substantial motions parallel to the solar surface, and their multi-component nature. More extensive studies near the limb will help in better establishing their characteristics and their origin.

\begin{acks} 
The authors gratefully acknowledge use of data from the IRIS data\-base. IRIS is a NASA small explorer mission developed and operated by LMSAL with mission operations executed at NASA Ames Research center and major contributions to downlink communications funded by ESA and the Norwegian Space Centre.  Additional data courtesy of NASA/SDO and the AIA and HMI science teams. Thanks are also due to Dr. A. Koukras for his comments.
\end{acks} 

\smallskip\noindent{\footnotesize{\bf Declaration on Conflicts of Interest}}
The authors declare that they have no conflicts of interest.

\smallskip\noindent{\footnotesize{\bf Data Availability Statement}}
The IRIS data used in this study are available at https://iris.lmsal.com/search/

\bibliographystyle{spr-mp-sola}
\bibliography{IRIS_Limb2}  

\newpage
\appendix
\section{Animation of Our \Si\ 1394\,\AA\ Data Set}\label{append}
\begin{figure}[h]
\begin{center}
\includegraphics[width=.85\textwidth]{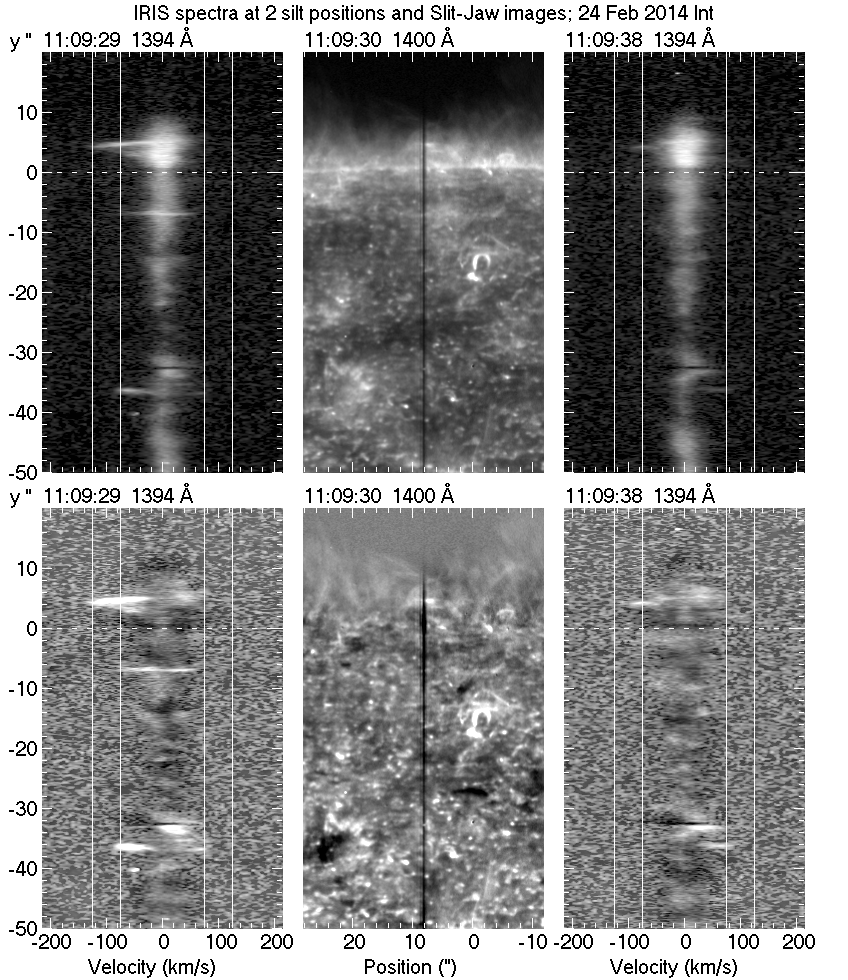}
\end{center}
\caption{A frame of the accompanying movie during EE1. {\it Top row\/}: Spectrum at slit position 1 ({\it left}), SJ image ({\it center}) and spectrum at slit position 2 ({\it right}). {\it Bottom row\/}: Same, with the time average subtracted. The dashed horizontal line marks the positiion of the continuum limb, the {\it vertical} lines mark the integration range of the position--time images discussed in Section~\ref{height}. Solar South is up.}
\label{movie}
\end{figure}

Figure~\ref{movie} shows a frame from the accompanying movie of \Si\ 1394\,\AA\ spectra and 1400\,\AA\ SJ images used in this work. The SJ images (middle column) were taken at the same time as the spectra at slit position 1 (left column). The spectra at slit position 2 (right column), located 0.9\arcsec\ west (left in the figure) of slit position 1, were taken at the same time as the 2796\,\AA\ SJs (not shown).

We note that, in addition to the explosive events, many other interesting features are visible in the SJs, such as the loop in the figure, another flat loop around 12:19 UT and several jets.

\end{article} 
\end{document}